\documentclass[reprint,amsmath,amssymb,aps,prd,longbibliography]{revtex4-1}
\usepackage{graphicx}
\usepackage{float} 
\usepackage[caption=false]{subfig} 
\usepackage{hyperref}
\usepackage{bm}

\newcommand{\appref}[1]{Appendix~\ref{#1}}
\begin{document}

\preprint{APS/123-QED}

\title{Identifying time dependence in network growth}

\author{Max Falkenberg$^{1,2,}$}
\email[Corresponding author: ]{max.falkenberg13@imperial.ac.uk}
\author{Jong-Hyeok Lee$^{3}$}
\author{Shun-ichi Amano$^{3}$}
\author{Ken-ichiro Ogawa$^{3}$}
\author{Kazuo Yano$^{4}$}
\author{Yoshihiro Miyake$^{3}$}
\author{Tim S. Evans$^{1,2}$}
\author{Kim Christensen$^{1,2}$}
\affiliation{$^1$Blackett Laboratory, Imperial College London, London SW7 2AZ, United Kingdom}
\affiliation{$^2$Centre for Complexity Science, Imperial College London, London SW7 2AZ, United Kingdom}
\affiliation{$^3$Department of Computer Science, Tokyo Institute of Technology, Yokohama, Kanagawa, Japan}
\affiliation{$^4$Center Research Laboratory, Hitachi Ltd., Kokubunji, Tokyo, Japan}

\date{\today}

\begin{abstract}
Identifying power-law scaling in real networks - indicative of preferential attachment - has proved controversial. Critics argue that measuring the temporal evolution of a network directly is better than measuring the degree distribution when looking for preferential attachment. However, many of the established methods do not account for any potential time-dependence in the attachment kernels of growing networks, or methods assume that node degree is the key observable determining network evolution. In this paper, we argue that these assumptions may lead to misleading conclusions about the evolution of growing networks.
We illustrate this by introducing a simple adaptation of the Barab{\'a}si-Albert model, the ``k2 model'', where new nodes attach to nodes in the existing network in proportion to the number of nodes one or two steps from the target node. The k2 model results in time dependent degree distributions and attachment kernels, despite initially appearing to grow as linear preferential attachment, and without the need to include explicit time dependence in key network parameters (such as the average out-degree). We show that similar effects are seen in several real world networks where constant network growth rules do not describe their evolution. This implies that measurements of specific degree distributions in real networks are also likely to change over time.
\end{abstract}

\pacs{Valid PACS appear here}
\maketitle

\section{Introduction\label{sec:intro}}

The study of complex networks has expanded rapidly over the past 20 years. Many real systems have been analyzed using networks with great success, showing many non-trivial properties \cite{vasiliauskaite2020}. Model networks have been defined to understand the origin and development of these properties from elementary principles. For instance, the Watts-Strogatz model generates networks with short average path lengths but high clustering coefficients, explaining the small world phenomenon \cite{watts1998}. Similarly, the Barab\'asi-Albert (BA) model, an undirected version of the Price model \cite{P76},  demonstrates that scale free (power-law) degree distributions in real networks can arise from a combination of growth and preferential attachment \cite{barabasi1999}. These models have given significant insight into the structure of real networks.
However, real systems almost never reflect the exact details of a model.

One of the most common features to study in a real network is the degree distribution \cite{broido2018}. The degree, $k$, of a node in a network is the number of direct connections a node has to other nodes in the network. The degree distribution, $P(k)$, is the probability distribution of the degree across all the nodes in the network. The degree distribution is said to be scale free (the exact definition is argued over) if it displays power-law scaling such that $P(k) \propto k^{-\kappa}$, where $\kappa$ is a positive constant, often found to be in the range $\kappa \in \{2,3\}$ for real networks \cite{newman2005}. Plotting $P(k)$ vs.\ $k$ on a log-log scale, a power-law distribution appears as a straight line with gradient $-\kappa$.

Since the late 1990s, many real networks have been reported as having scale free, or nearly scale free degree distributions. This includes web-page links on the internet \cite{albert1999}, citation networks \cite{redner1998}, the co-occurrence of words in language \cite{cancho2001}, sexual contact \cite{liljeros2001}, social networks \cite{java2007}, and others. Identifying these networks as scale free has important consequences: (1) it gives a potential mechanistic understanding of the origin and development of these networks, notably that the network evolves according to preferential attachment, and (2) it suggests these networks have a set of important properties associated with scale free networks. These properties include the presence of hub nodes which have degree much larger than the network average, very small network diameters \cite{cohen2003}, and resistance to errors but vulnerability to targeted attack \cite{crucitti2003}.

Although the scale free paradigm has become a hallmark of complex networks research, identifying scale free behavior in real networks is still very controversial \cite{holme2019}.

Significant effort has gone towards developing appropriate statistical techniques to assess whether networks at a fixed point in time are scale free, most notably in \cite{clauset2009,broido2018}. Given the difficulty in distinguishing a power-law distribution, $P(k) \propto k^{-\kappa}$, from similar distributions such as the log-normal, $P(k) \propto k^{-1} \text{exp}[-\frac{(\text{ln} k-\mu)^2}{2\sigma^2}]$, or stretched exponential, $P(k) \propto k^{\tau - 1}\text{e}^{-\lambda k ^{\tau}}$,
these statistical techniques are clearly important for understanding the statistical properties of network degree distributions. Applying such techniques to a large set of real world networks, a recent study found that true scale free networks are rare, representing only about 4\% of all networks \cite{broido2018}. These results are broadly in line with a number of similar criticisms of the scale free paradigm \cite{RFC08,lima2009,willinger2009,golosovsky2017,stumpf2012,Sendina2016,kang2020}. However, despite broad support for these criticisms, many others in the networks community are still strong believers that most complex networks exhibit preferential attachment \cite{goh2002,House2015,barabasi2016,voitalov2019}. Among these individuals, many have taken issue with the methods to process the data in \cite{broido2018} and/or the strictness of the scale free definition \cite{barabasi2016,holme2019,voitalov2019,gerlach2019}. Arguing that scale free networks are only well defined in the infinite system size limit, looser definitions suggest that scale free networks are in fact not rare at all \cite{voitalov2019}. However, it can be argued that such a loosening will naturally result in a larger number of positive identifications, and that using weakened criteria for scale freeness defeats the aim of using a statistically rigorous approach. Clearly, the issue of which approach is best when analyzing network degree distributions is yet to be fully resolved.

There is a third camp who argue that ``knowledge of whether or not a distribution is heavy-tailed is far more important than whether it can be fit using a power-law'' \cite{stumpf2012}. However, great care must be take with such an approach in a context dependent manner. For instance, in the case of epidemic spreading, two networks may both be fat-tailed with similar degree-distributions, yet exhibit very different epidemic mixing patterns due to differences in network assortativity \cite{kiss2008,piraveenan2012}.

What all these approaches have in common is that they analyze the degree distribution of a network at a fixed point in time. If such an analysis is to give insight into the mechanistic origin and evolution of a network, it would be prudent to ask whether the degree distribution is representative of the network in general during its evolution, or only for a brief period of time? Without an answer to this question, inferring the past and future evolution of a network based on the current form of its degree distribution may give misleading results.

A prominent example of a theoretical network model where the observed degree distribution appears to change over time is super-linear preferential attachment, where new nodes attach to existing nodes proportionally to their degree to a power greater than 1 \cite{krapivsky2000}. In the long time limit, a gelation phenomenon is observed where almost all nodes connect to a single hub node forming a star-like network. However, \citeauthor{krapivsky2008} \cite{krapivsky2008} showed that super-linear attachment has significant pre-asymptotic regimes where the degree distribution appears to be approximately scale free.

Given the difficulty of directly identifying preferential attachment from static degree distributions, proponents of the scale free paradigm have argued that preferential attachment can be identified directly from dynamical network data (if available) \cite{barabasi2016}. Numerous approaches have been introduced over the years, using a variety of different assumptions \cite{newman2001,jeong2003,massen2007,gomez2011,sheridan2012,kunegis2013,pham2015}. Most commonly, methods assume that the preferential attachment kernel follows a functional form, $\Pi (k) \propto k^{\gamma}$, and primarily focus on estimating the exponent $\gamma$ - such methods will naturally assume that the preferential attachment kernel of a network is {\it time independent}.

As an alternative approach, non-parametric methods have been proposed that do not assume a functional form. The first of these methods by \citeauthor{jeong2003} \cite{jeong2003} infers the form of the attachment kernel by constructing a histogram of the degree of nodes to which new edges attach over a short observation window. However, there is no clear guide as to how to choose the start of the observation window and how long it must be - too short and the result is very noisy, too long and the result is subject to bias \cite{pham2015}. The method by \citeauthor{newman2001} \cite{newman2001} avoids this problem by constructing multiple histograms over different observation windows and computes the attachment kernel by taking a weighted average over the different histograms. Although this method avoids the issue of how to choose your observation window, this approach seems to underestimate the attachment kernel at large degrees \cite{herdagdelen2007}, an issue since corrected by \citeauthor{pham2015} \cite{pham2015}.

For networks in which the attachment kernel is time independent, the corrected Newman method proposed in \cite{pham2015} gives an excellent fit to data. However, it is still not clearly established whether the assumption of time independence is valid for real networks, and in some cases (such as citation networks) it is known to be false \cite{leskovec2005}. Similarly, the probability of attaching to a node may be a function of a variable other than the degree. However, how to correctly identify which feature of a node determines its attractiveness is not clear.

It is often argued that accurately calculating the attachment kernel of a growing network is important because it can help to predict the future evolution of a network \cite{kunegis2013}. For instance, in the case of non-linear preferential attachment, where the attachment kernel is given by $\Pi (k) \propto k^{\gamma}$ with positive constant $\gamma$, it is known that for $0 < \gamma < 1$, the limiting degree distribution is a stretched exponential, whereas for $\gamma > 1$, the degree distribution displays a gelation phenomenon where a single dominant hub connects to almost all other nodes in the network \cite{krapivsky2000}. In between, $\gamma = 1$ corresponds to traditional linear preferential attachment where the degree distribution displays power-law scaling. Hence, if we can estimate the value of $\gamma$ for the attachment kernel of a real network, this can be used to predict its future evolution.

Predictions regarding the future evolution of networks, explanations of the historical development of networks, and investigations into whether preferential attachment underlies the evolution of networks, based on measured attachment kernels, are widespread in the literature. These include studies on citation networks \cite{rondapupo2018,sheridan2018}, protein networks \cite{eisenberg2003}, the bitcoin network \cite{kondor2014}, common words in the English language \cite{perc2012}, social dynamics in online games \cite{szell2010}, actor networks \cite{jeong2003}, and more.

The majority of these studies make three assumptions: (1) that the degree of a node is the key feature determining a node's attractiveness, (2) that the attachment kernel can be approximated by $\Pi (k) \propto k^{\gamma}$, and (3) that the measured attachment kernel is either time independent, or that the time dependence is largely unimportant. For instance, looking at four different periods in the evolution of the {\it American Physical Society} (APS) citation network, and using the node degree (citation count) as the key variable of interest, \citeauthor{sheridan2018} found that the exponent $\gamma$ ranges from 0.94 to 1.06 \cite{sheridan2018}. The authors assert that this implies that the attachment probabilities in the APS citation network are at least approximately time independent. However, as noted, $\gamma < 1$ would imply that the APS citation network's degree distribution approaches a stretched exponential, whereas $\gamma > 1$ would result in a gelation effect. Since both $\gamma < 1$ and $\gamma >1$ were observed from the data, what does this imply for the future evolution of the network?

The aim of this paper is to illustrate the risks of assuming time independence in the rules governing the evolution of growing networks, and the risk of assuming that the node degree determines node attractiveness. We will do this by introducing the ``k2 model'', a simple variant of the Barab{\'a}si-Albert model where new nodes do not attach to existing nodes proportionally to the number of direct neighbors a node has, but rather proportionally to the number of nodes within a distance two of the target node. This simple rule is rooted in the idea that well connected neighbors are preferable to poorly connected neighbors. The rule puts a particular focus on the role of nearest neighbor correlations in network growth. Such mechanisms of mutual benefit may be relevant to collaboration \cite{Li2019}, or citation \cite{vazquez2000} networks. The mechanism may also have indirect relevance to node copying processes \cite{lambiotte2016,bhat2016,krapivsky2005}. Similar ideas have been explored in \cite{dangalchev2004,Wang2018,Topirceanu2018,amano2018}.

Although this simple rule has no explicit time dependence (i.e.\ time dependence is not in built by including a time dependent parameter, e.g., the average out-degree), the correlations that form between neighboring nodes result in an implicit time dependence in the attachment kernel. Consequently, the resulting network does not demonstrate any of the simple scaling observed in traditional network models. This is despite an extended initial transient phase during which the network appears to grow according to linear preferential attachment. We support this argument with an analytical treatment demonstrating that assumptions of simple scaling in the k2 model are not robust.

The arguments we illustrate with the k2 model are highly relevant to real networks. By calculating the ratio of network attachment kernels over different time periods, we show that over short timescales, assumptions of time independence for real networks are relatively well justified. However, over longer time periods, the relative attachment kernels calculated show clear time dependence, displaying a diversity of patterns. While the overall effect may be small in some cases (such as for the Flickr friendship network or the English Wikipedia hyperlink network \cite{mislove2009}), we argue that, at a minimum, practitioners should test the degree of time dependence in their data before making predictions about the future or past development of a network.

\section{Methods}
\subsection{Model Definition}

The k2 model is defined as a simple, undirected network. The model is initialized with a small connected network of $m_{0}$ nodes. Each time-step, a new node is created with $m \leq m_0$ new edges. The $m$ edges are connected to the new node and target nodes from the network. Each target node is chosen with probability proportional to the number of neighbors which are one or two steps away, $k_{i}^{(2)}$, from the target node, $i$. We refer to $k_{i}^{(2)}$ as the second degree of node $i$, see \appref{ap:maths} for a formal definition. The attachment probability is identical to the BA model with the exception that the BA model attaches proportionally to the number of nodes one step away, $k_{i}^{(1)}$, from the target node, $i$. We refer to $k_{i}^{(1)}$ as the first degree, or just the degree, of node $i$.
Computationally, we prevent multiple edges being formed between two nodes by selecting $m$ unique target nodes.

For clarity, whenever notation is presented with a subscript $i$ or $j$, for instance $k_{i}^{(1)}$ or $k_{j}^{(2)}$, the focus is on the value of that variable for the particular node $i$ or $j$. When the subscript is omitted, for instance $k^{(1)}$ or $k^{(2)}$, the focus is on all nodes with the same specific value of the variable in question. We use $k$ and $k^{(1)}$ interchangeably where appropriate.

\vspace{3pt}
\begin{figure}[h!]
\centering
\includegraphics[width=\linewidth]{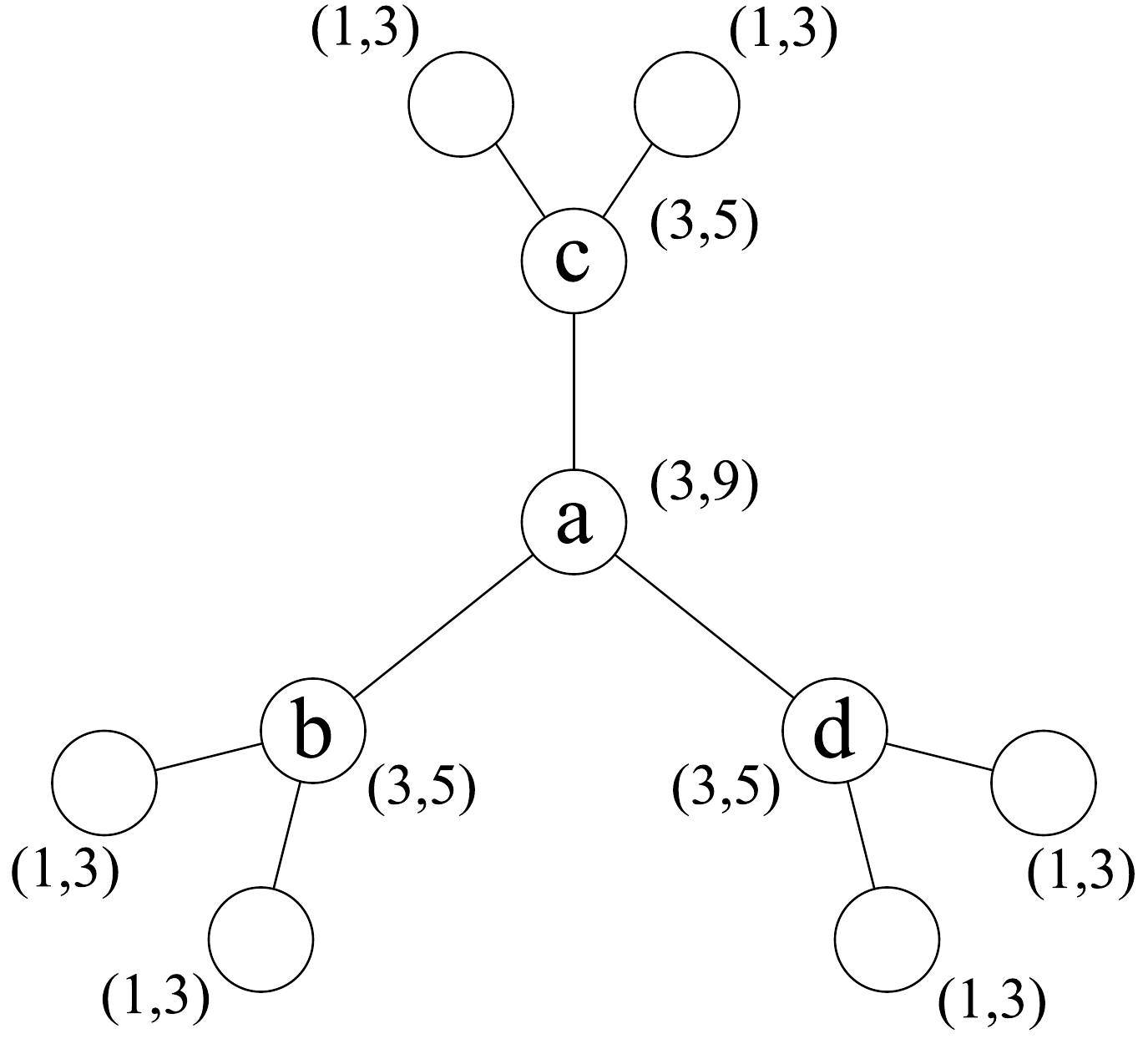}
\caption{A sketch of a simple tree network with four labeled nodes. The values in the brackets correspond to the first and second degree, $(k_i^{(1)},k_i^{(2)})$, of each node. The four labeled nodes have the same degree, $k_i^{(1)} = 3$, implying equal importance in the BA model, but different second degrees, implying unequal importance in the k2 model.}
\label{fig:fig1}
\end{figure}
\vspace{3pt}

\begin{figure*}
\centering
\includegraphics[width=\linewidth]{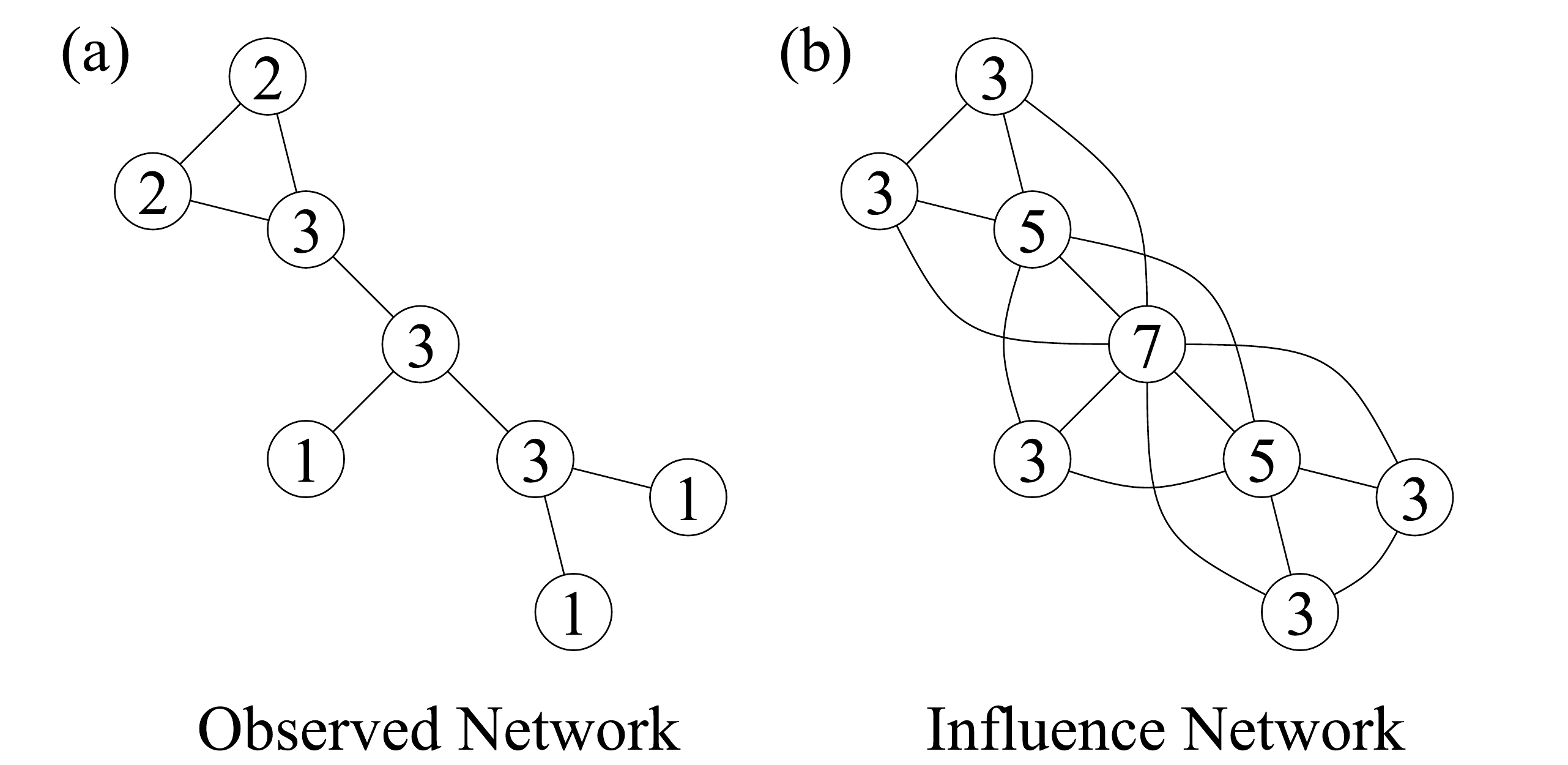}
\caption{An illustration of how the k2 model can be thought of as generating two distinct networks. (a) The \textit{observed} network is the network of nodes which have a direct connection to each other. In this network, the degree does not account for the importance of next-nearest neighbors. (b) The \textit{influence} network where two nodes are connected if they are nearest or next-nearest neighbors in the observed network.}
\label{fig:fig2}
\end{figure*}

Figure~\ref{fig:fig1} illustrates the motivation for the k2 model. In the BA model, a node's importance is proportional to the number of nodes connected to it, i.e.\ the first degree. However, there is no consideration for whether these connected nodes are important or not. A node with three isolated neighbors is considered equally important to a node neighboring three hubs. This is in conflict with many real world scenarios, for instance in academic collaboration networks, where it is known that junior researchers working under top scientists are those most likely to be successful and reach tenure in their careers \cite{Li2019}. In the k2 model, this effect is accounted for, allowing nodes to benefit from connecting to hub nodes and giving them the opportunity to become hubs themselves.

The principle of weighting neighbor importance reflects the role of friends-of-friends in social network theory \cite{granovetter1977,B92a}, and is the foundation for widely used network centrality measures built on self-consistent equations, such as Katz centrality \cite{K53,newman2018} or PageRank \cite{BP98,newman2018}. 

Mathematically, we define the attachment kernel, $\Pi$, as the function specifying the probability of attaching to a specific node in the network. In the BA model, $\Pi^{(BA)} \propto k^{(1)}$, whereas in the k2 model, $\Pi^{(k2)} \propto k^{(2)}$. In the case of the k2 model, we can write the normalized form of the attachment kernel as
\begin{equation}
    \Pi^{(k2)}_{i} = \frac{k^{(2)}_{i}}{\sum_{j=1}^{N} k_{j}^{(2)}} \approx \frac{k^{(2)}_{i}}{\sum_{j=1}^{N} (k_{j}^{(1)})^{2}},
    \label{eq:kernel}
\end{equation}
where for $m=1$, the approximation is an equality. For $m>1$ the approximation holds as long as the number of non-unique second degree neighbors is small, see \appref{ap:maths}. By splitting the numerator of the attachment kernel into the contribution of the first degree neighbors to node $i$, $k_{i}^{(1)}$, and the contribution of the next-nearest neighbors, $k_{i}^{(2)} - k_{i}^{(1)}$, Eq.~\eqref{eq:kernel} can be rewritten as
\begin{equation}
    \Pi_{i}^{(k2)} \approx \frac{k_{i}^{(1)} + \sum_{\alpha = 1}^{k_{i}^{(1)}} (k_{i\alpha}^{(1)} - 1)}{\sum_{j=1}^{N} (k_{j}^{(1)})^{2}},
    \label{eq:kernel_simple}
\end{equation}
which is a function of the first neighbor degree only, where we have used
\begin{equation}
    k_{i}^{(2)} = \sum_{\alpha = 1}^{k_{i}^{(1)}} k_{i\alpha}^{(1)}.
    \label{eq:k2def}
\end{equation}
Here, $\alpha$ labels the $k_{i}^{(1)}$ unique first neighbors of node $i$, and $k_{i\alpha}^{(1)}$ is the first degree of node $\alpha$, connected to node $i$. In Eq.~\eqref{eq:kernel_simple}, the first term indicates the contribution to the attachment kernel from the direct neighbors of node $i$, and the second term indicates the contribution from next-nearest neighbors to node $i$.

Conceptually, we can think of the k2 model as involving two separate networks. In the \textit{observed} network, each node represents an agent, and an edge between two nodes represents a direct, first degree relationship between the two nodes. However, new nodes do not connect to a target node according to the node's direct connections, but rather according to the number of nodes within distance two of the target. These nodes are within the sphere of influence of the target node. Hence, we define the \textit{influence} network, in which an edge between any two nodes signifies that the nodes are within each other's sphere of influence, i.e., two connected nodes are neighbors, or next-nearest neighbors in the observed network, see Fig.~\ref{fig:fig2}.

The influence network has similarities to the node copying mechanism studied in \cite{lambiotte2016,bhat2016}, based on earlier models in \cite{krapivsky2005}, with relevance to social network formation \cite{granovetter1977,toivonen2009}, citation networks \cite{GAE15,vazquez2000}, evolution \cite{ohno2013}, and protein interaction networks \cite{ispolatov2005,kim2002}. Although the k2 model is not designed to model such systems explicitly, it may be useful for understanding the role of neighbor-neighbor correlations in the growth of such networks.

In the influence network, new nodes connect to a target node proportionally to the node's degree, $k^{(2)}$. The new node then copies a fraction of the nodes attached to the initial target node, and forms additional edges to these copied neighbors. The copied neighbors correspond to those which are directly connected to the target node in the observed network. In the node copying model, new nodes select a target node at random and then copy a fraction of the target node's neighbors. As opposed to the k2 model, the copied neighbors are selected at random with probability $p$. In this respect, the node copying model where the original target node is chosen preferentially could represent a mean-field version of the k2 model, where we neglect correlations between neighboring nodes.

\subsection{Measuring Time Dependence of Preferential Attachment}

To understand how the attachment kernel of a network changes over time, it is helpful to consider relative attachment probabilities as opposed to absolute attachment probabilities. In general we can write an arbitrary attachment kernel, which is a function of the node degree only, as
\begin{equation}
    \Pi(k;t) = \frac{f(k)}{\sum_{j=1}^{N(t)} f(k_j(t))}
    \label{eq:attachment_general}
\end{equation}
with an arbitrary preference function $f$. The summation is over all nodes in the network at time $t$. The function $f$ is \textit{time independent}, however, as the network grows and more nodes are added, $N(t)$ in the denominator changes, and hence, the denominator is \textit{time dependent}. Note, $f(k_i(t))$ for a specific node $i$ is \textit{time dependent}, since the degree of a specific node evolves over time. We define the relative attachment kernel as
\begin{equation}
    \phi_t(k,k') = \frac{\Pi(k;t)}{\Pi(k';t)} = \frac{f(k)}{f(k')}.
\end{equation}
As opposed to $\Pi(k;t)$, the relative attachment kernel has no dependence on the network as a whole, but rather, is a function of the degree $k$ and $k'$ only. As a result, we can express the time independence of the attachment kernel as
\begin{equation}
    \frac{\mathrm{d} \phi_t(k,k')}{\mathrm{d} t} = 0.
\label{eq:timeinvariance}
\end{equation}
For convenience, in the following we will consider $\phi_t(k,1)$, i.e., the attachment probability of connecting to a node with degree $k$ relative to a node with degree $k' = 1$. By definition, $\phi_t(k,k) = 1$.

Consider the relative attachment kernel calculated at time $t$, written as $\phi_{t}(k,1)$, and at time $s$, $\phi_{s}(k,1)$. If Eq.~\eqref{eq:timeinvariance} holds, then $\phi_{t}(k,1) = \phi_{s}(k,1)$. For a real network, it is likely that there will be small deviations from this ideal case. Hence, we can plot the ratio $\phi_{t}(k,1)/\phi_{s}(k,1)$ against degree $k$ to gauge the extent of the time dependence across a specific time interval. This ratio is only well defined for networks which contain nodes with degree $k$ at both times $t$ and $s$.

The BA model is a simple case where Eq.~\eqref{eq:timeinvariance} should hold, with $\phi_t(k,1) = k$ for all $t$. Likewise, for non-linear preferential attachment, $\phi_t (k,1) \propto k^{\gamma}$ with positive constant $\gamma$. In the case of the k2 model, Eq.~\eqref{eq:timeinvariance} does not hold, due to the second term in Eq.~\eqref{eq:kernel_simple}. The preference function in the k2 model is not a function of the degree of a node, but the second degree, $k^{(2)}$. Clearly the second degree is related to the first degree, and, when analyzing the k2 model, one could mistakenly believe that the node degree, $k^{(1)}$ is the quantity determining network growth. However, although this appears approximately true at first, over time, the relation between the first and second degree changes, $\left \langle k^{(2)}(t)\right\rangle\not\propto\left \langle k^{(1)}(t)\right\rangle$. In other words, although the attachment kernel is not explicitly time dependent (e.g., we have not included an explicit aging mechanism), the local network structure, which determines a node's second degree, is time dependent.

This point cannot be overstated; while the k2 model clearly breaks the assumptions outlined above, it does so in a way that, without prior knowledge of the model rules, is wholly non-obvious. As we will outline, if sufficient care is not taken, these assumptions risk misleading or incorrect predictions about a networks past or future evolution.

\section{Results\label{sec:results}}
\subsection{Simulation Results}
\begin{figure}
\centering
    \subfloat{\includegraphics[width = \linewidth]{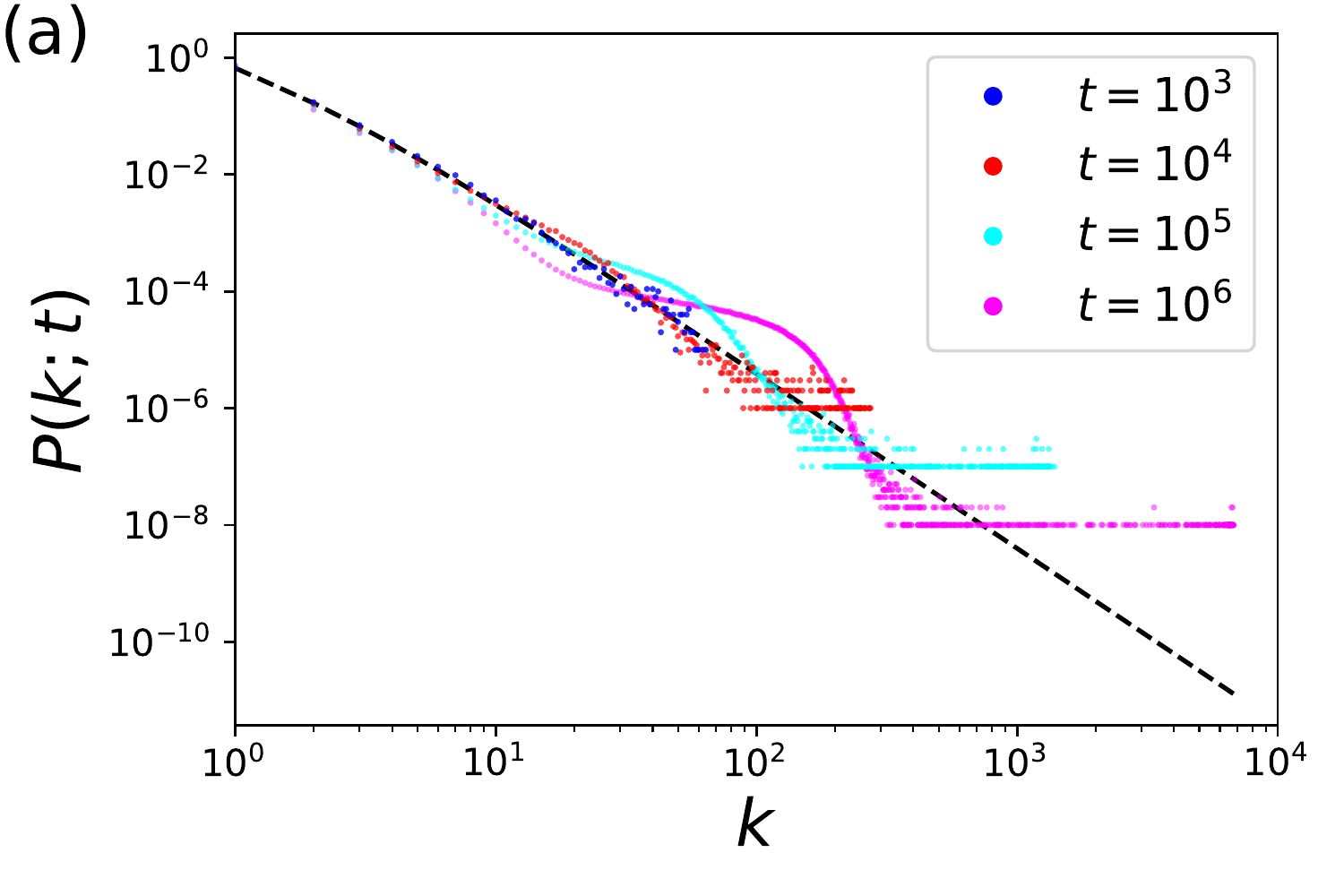}}
    \newline
    \subfloat{\includegraphics[width = \linewidth]{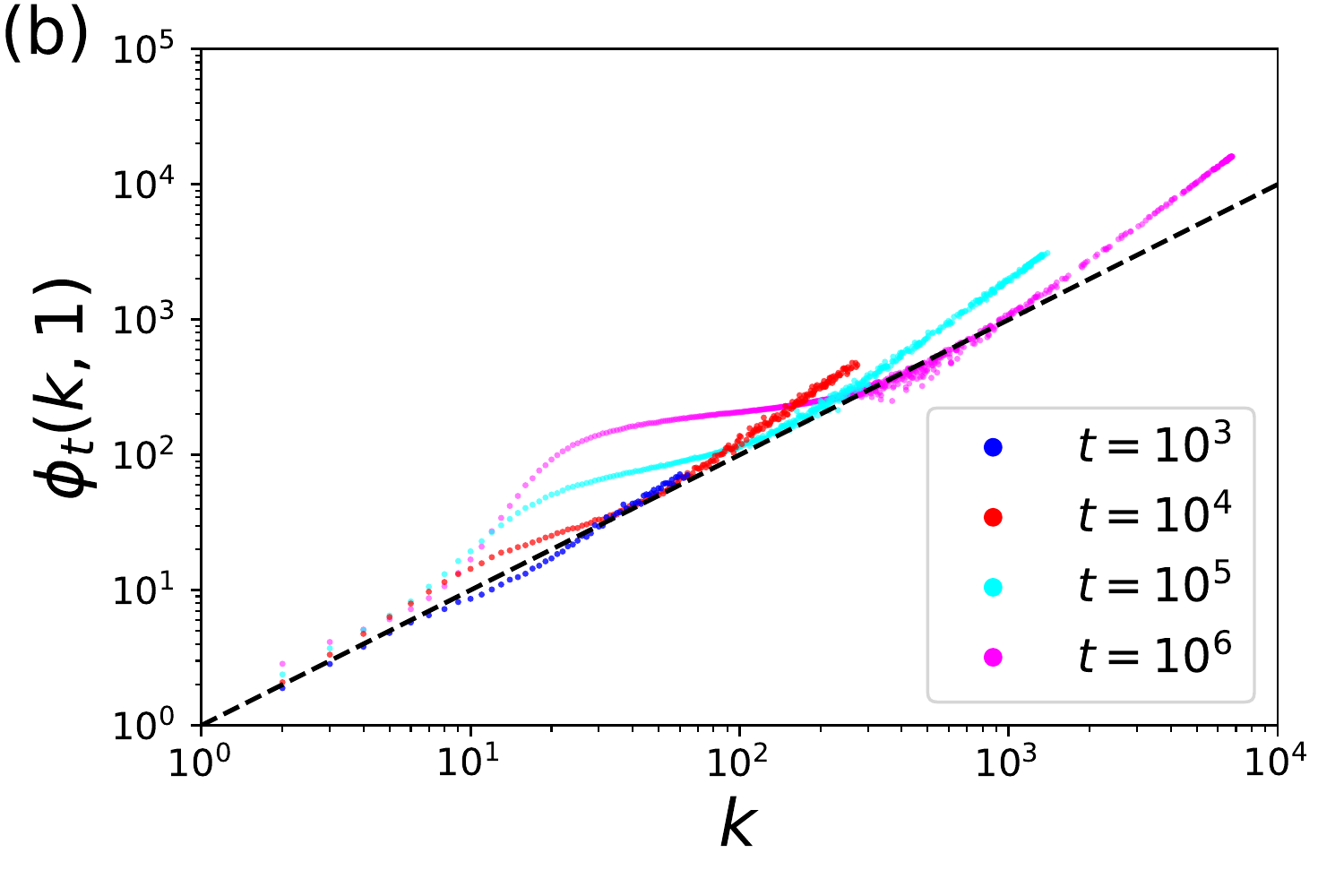}}
    \caption{The (a) degree distribution and (b) relative attachment kernel for the k2 model with $m=1$. The dashed lines show the expected scaling for the BA model. Early in the growth of the k2 model, the evolution of the network is largely indistinguishable from the BA model. As the network grows, both the degree distribution and relative attachment kernel deviate significantly from the simple BA model scaling.}
    \label{fig:k2_degree}
\end{figure}
\begin{figure*}
    \centering
    \subfloat{\includegraphics[width =0.5 \linewidth]{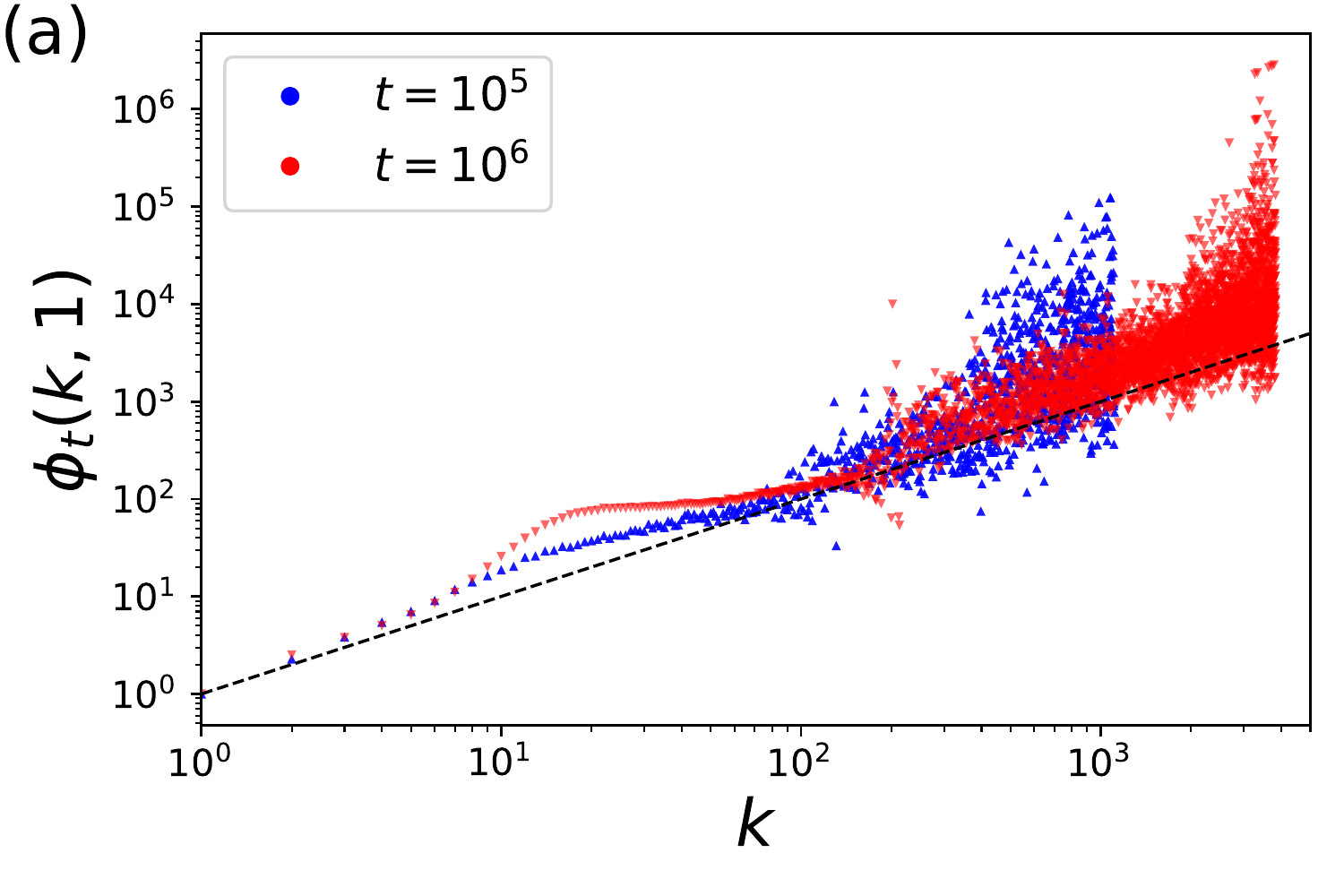}}
    \subfloat{\includegraphics[width = 0.5 \linewidth]{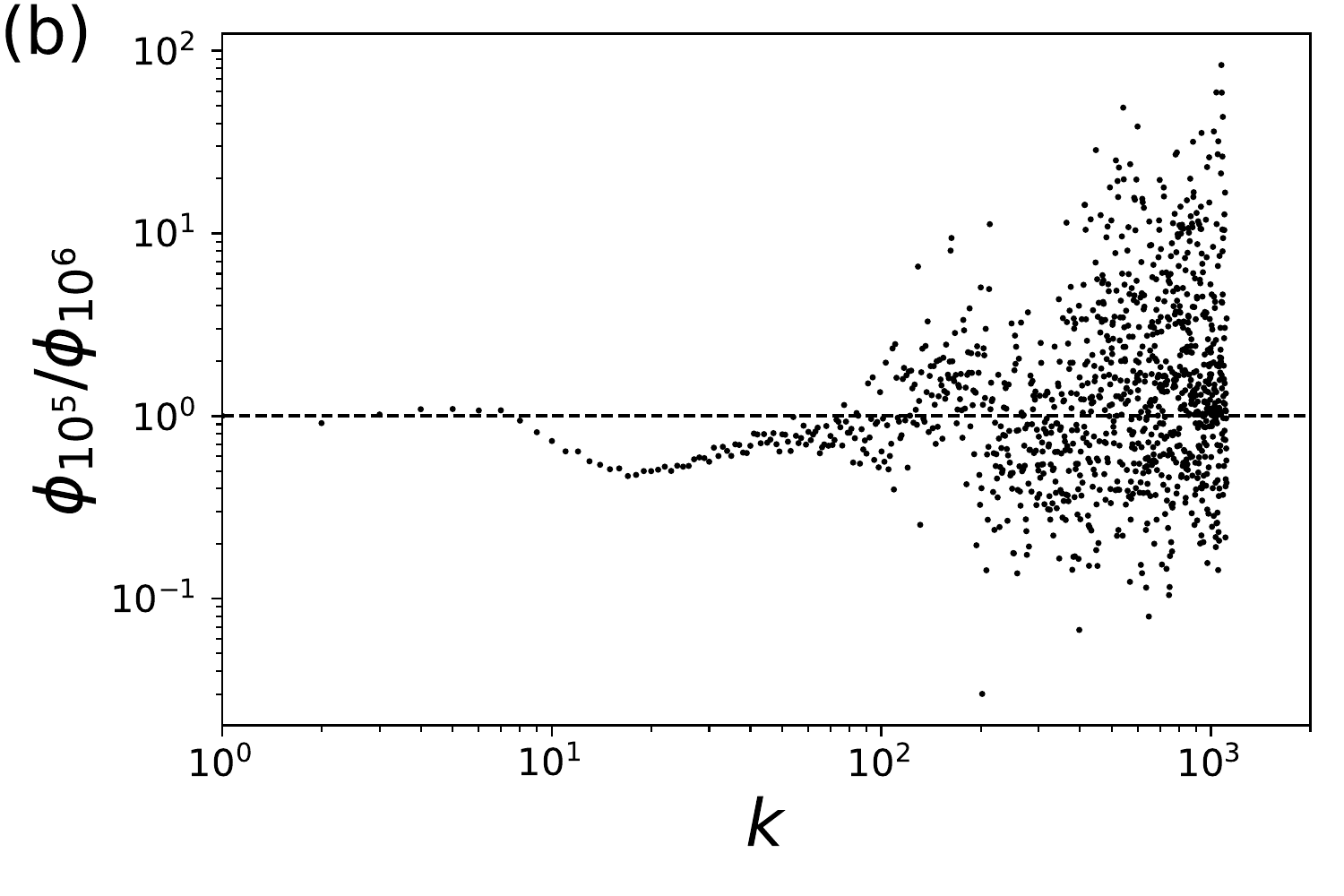}}
    \newline
    \subfloat{\includegraphics[width = 0.5 \linewidth]{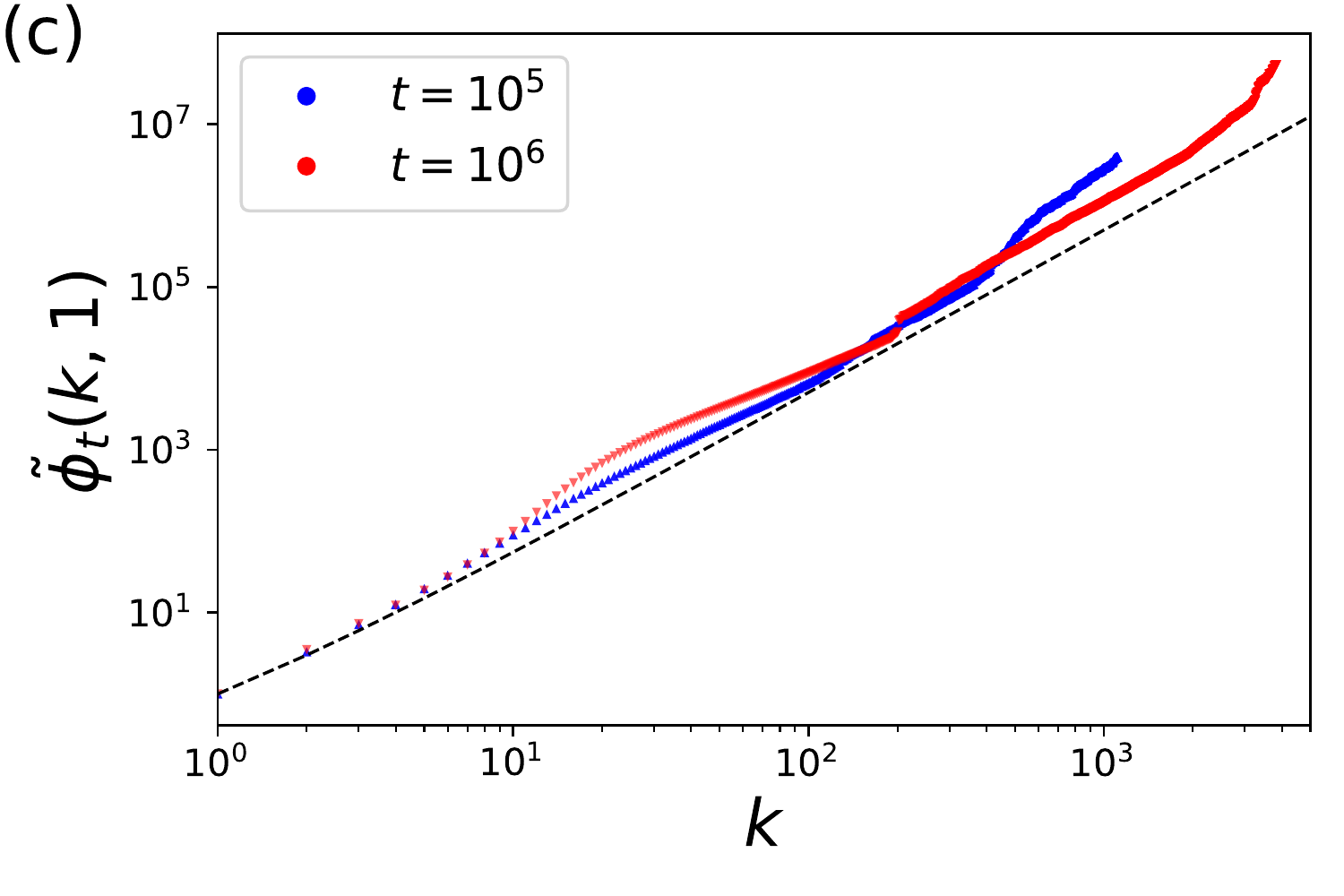}}
    \subfloat{\includegraphics[width = 0.5 \linewidth]{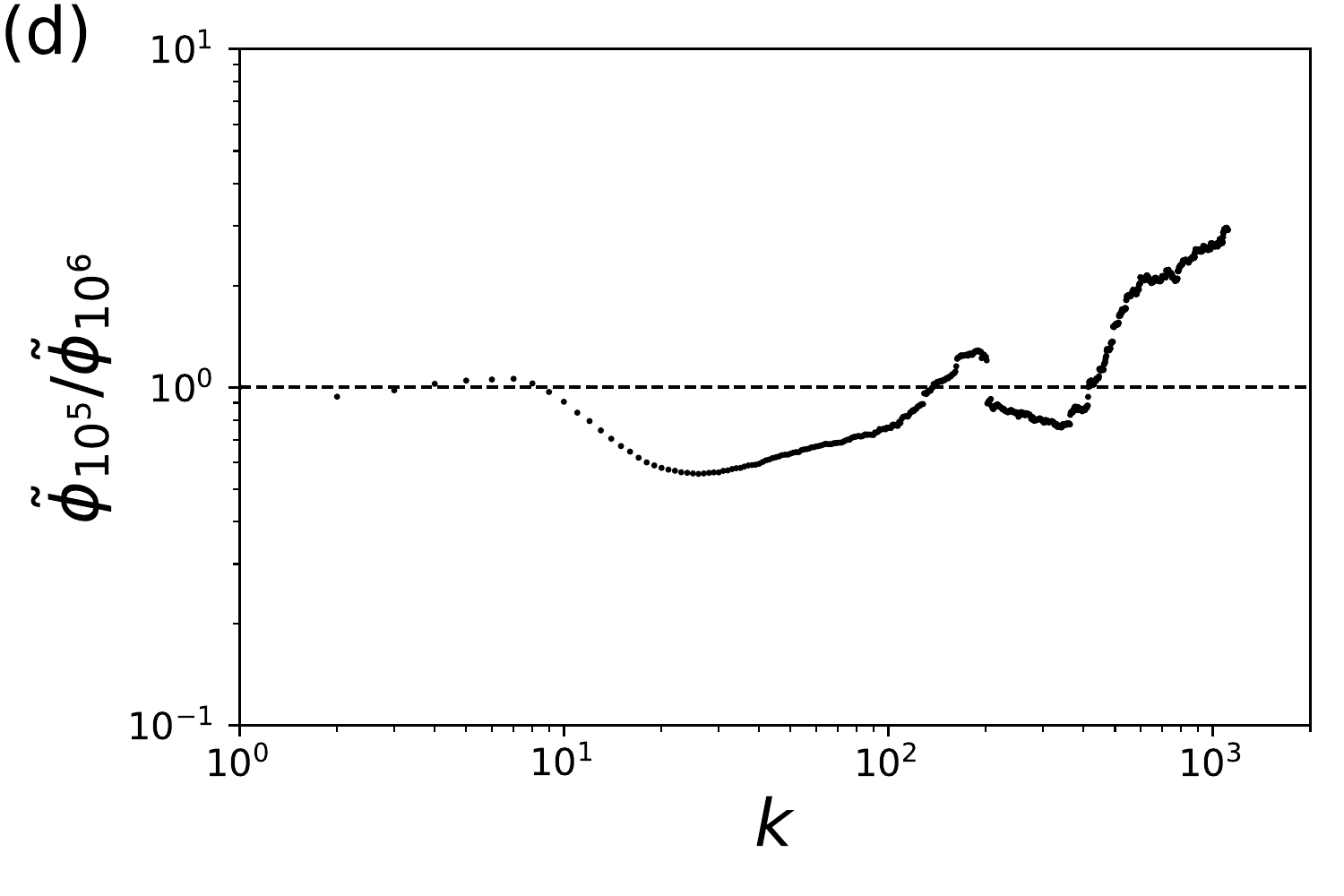}}
    \caption{An illustration that the relative attachment kernel in the k2 model is \textit{time dependent}. (a) The relative attachment kernel, $\phi_{t}(k,1)$, for the k2 model with $m=1$ as calculated using the corrected Newman method. (b) The ratio of the calculated relative attachment kernels. (c) The cumulative sum and (d) corresponding ratio of the relative attachment kernels shown in (a). The dashed lines indicate the prediction for the BA model in (a) and (c), and the ratio expected for any time independent relative attachment kernel, which is a function of node degree only, in (b) and (d).}
    \label{fig:k2_phi_calc}
\end{figure*}
\begin{figure*}
    \centering
    \subfloat{\includegraphics[width =0.5 \linewidth]{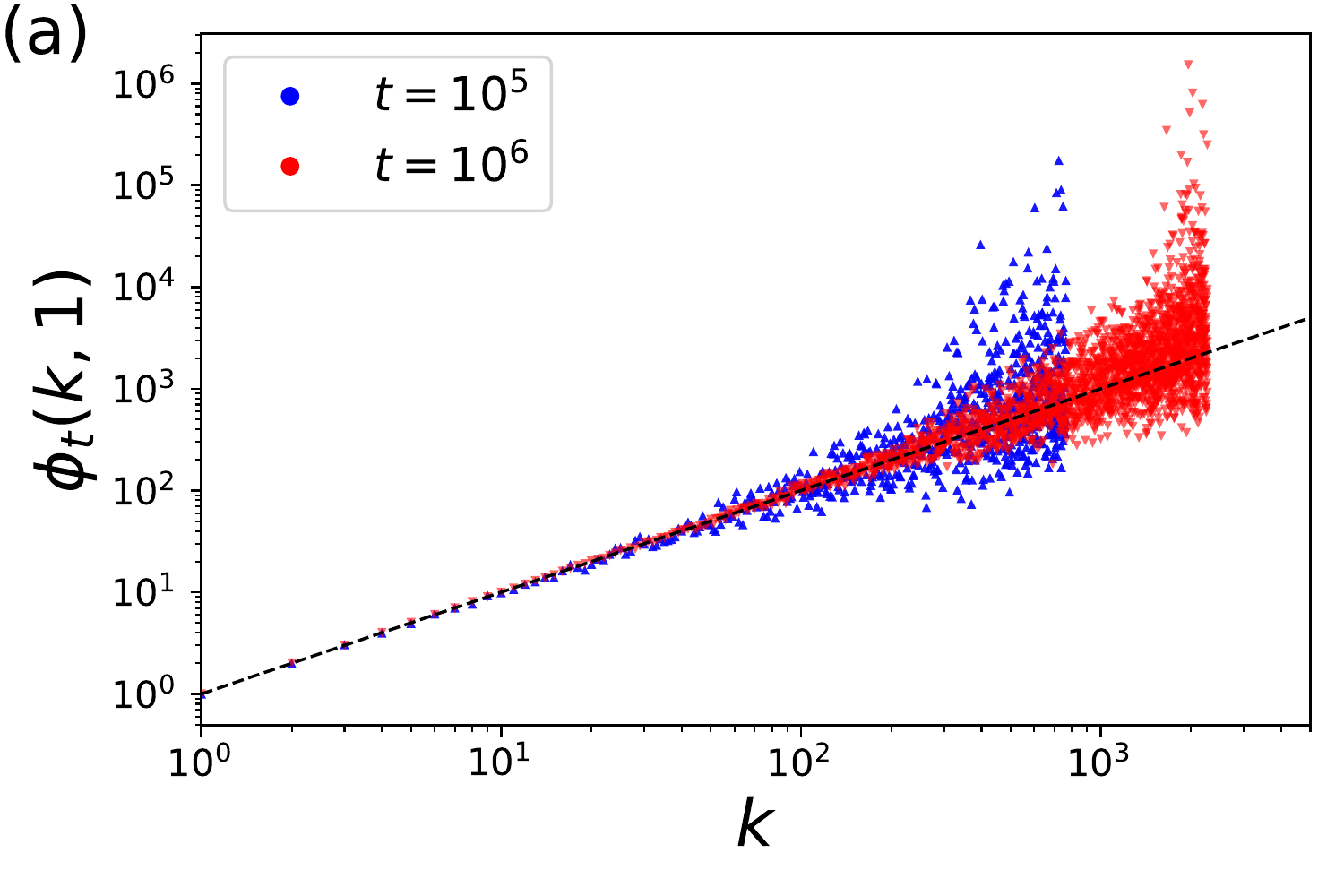}}
    \subfloat{\includegraphics[width = 0.5 \linewidth]{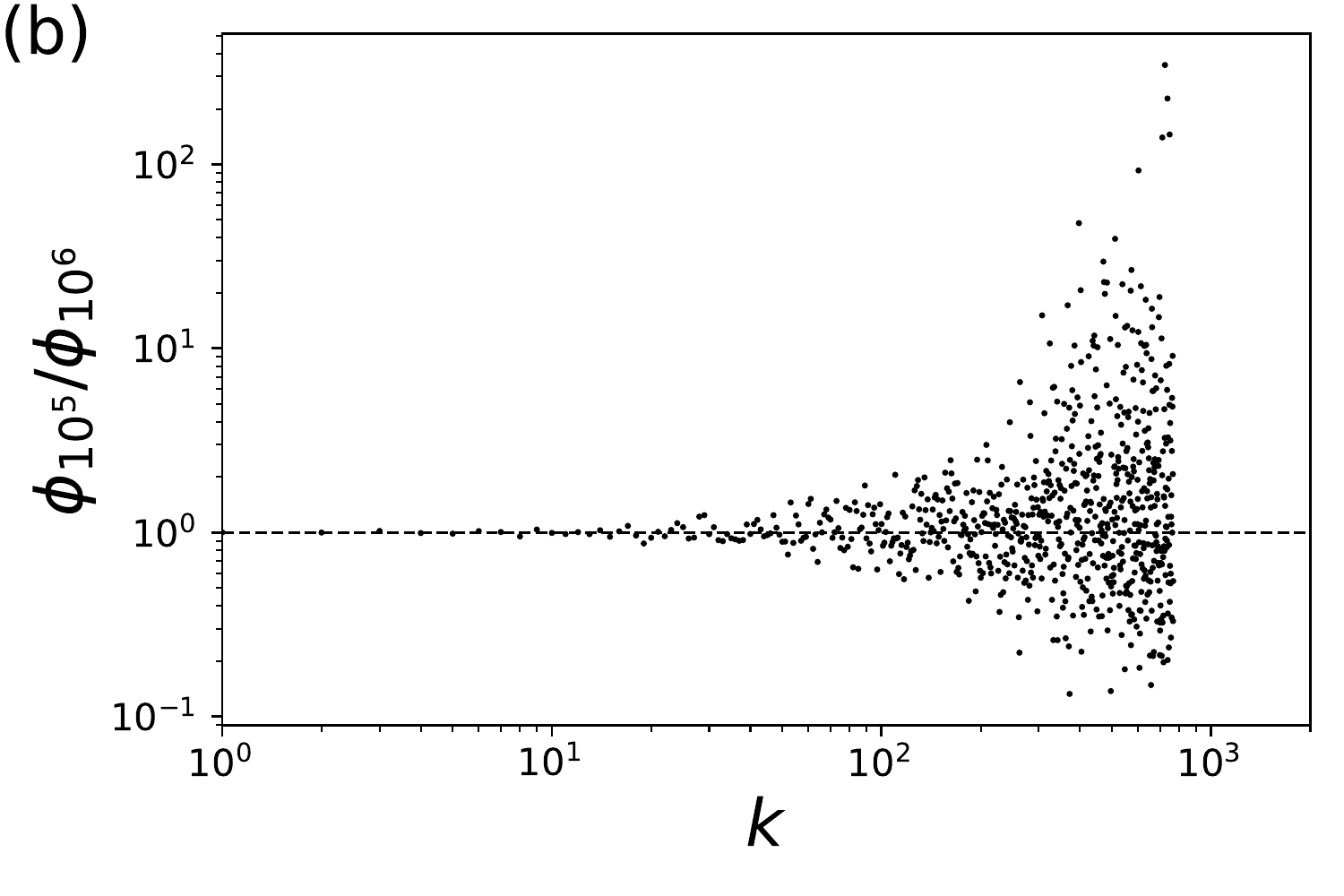}}
    \newline
    \subfloat{\includegraphics[width = 0.5 \linewidth]{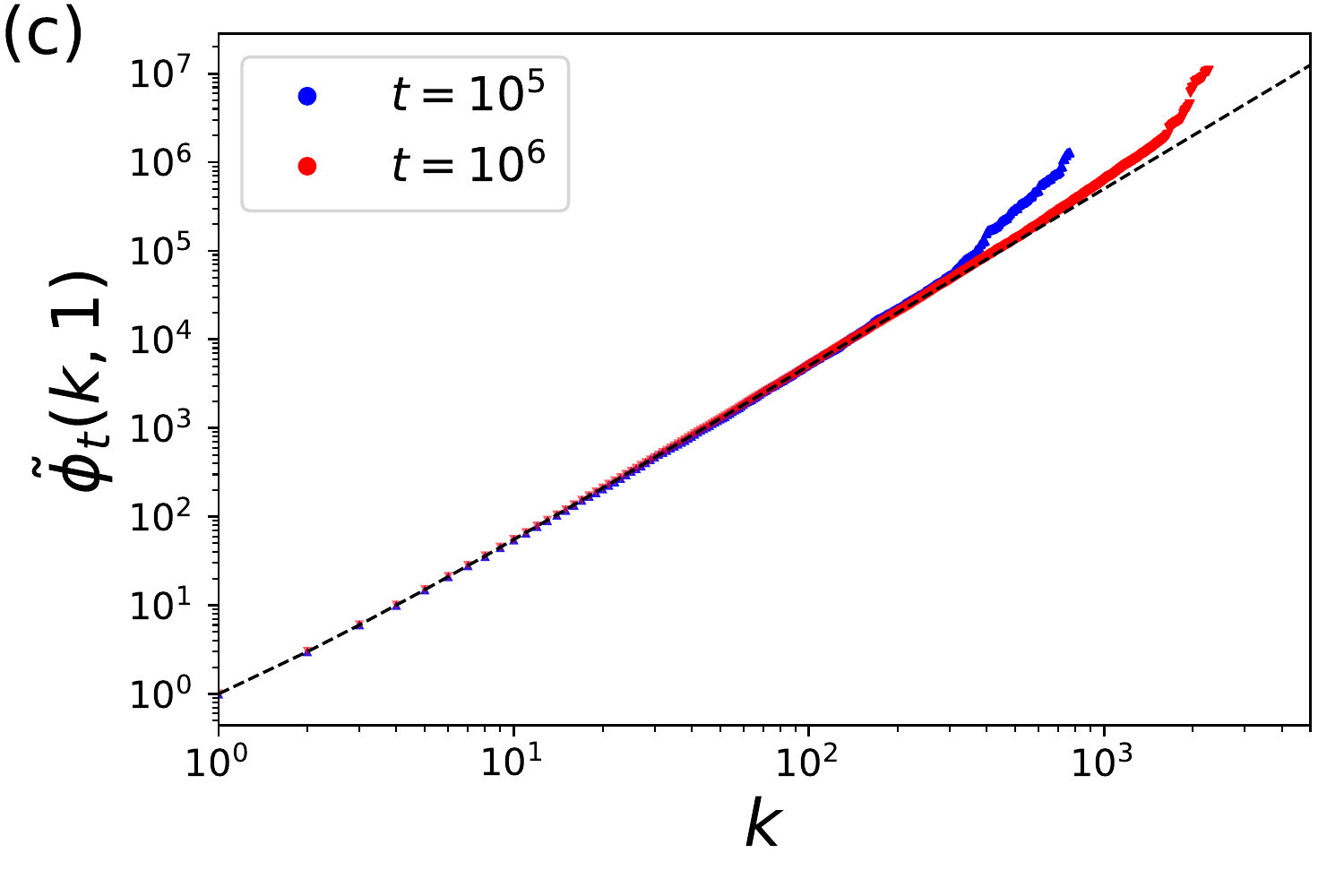}}
    \subfloat{\includegraphics[width = 0.5 \linewidth]{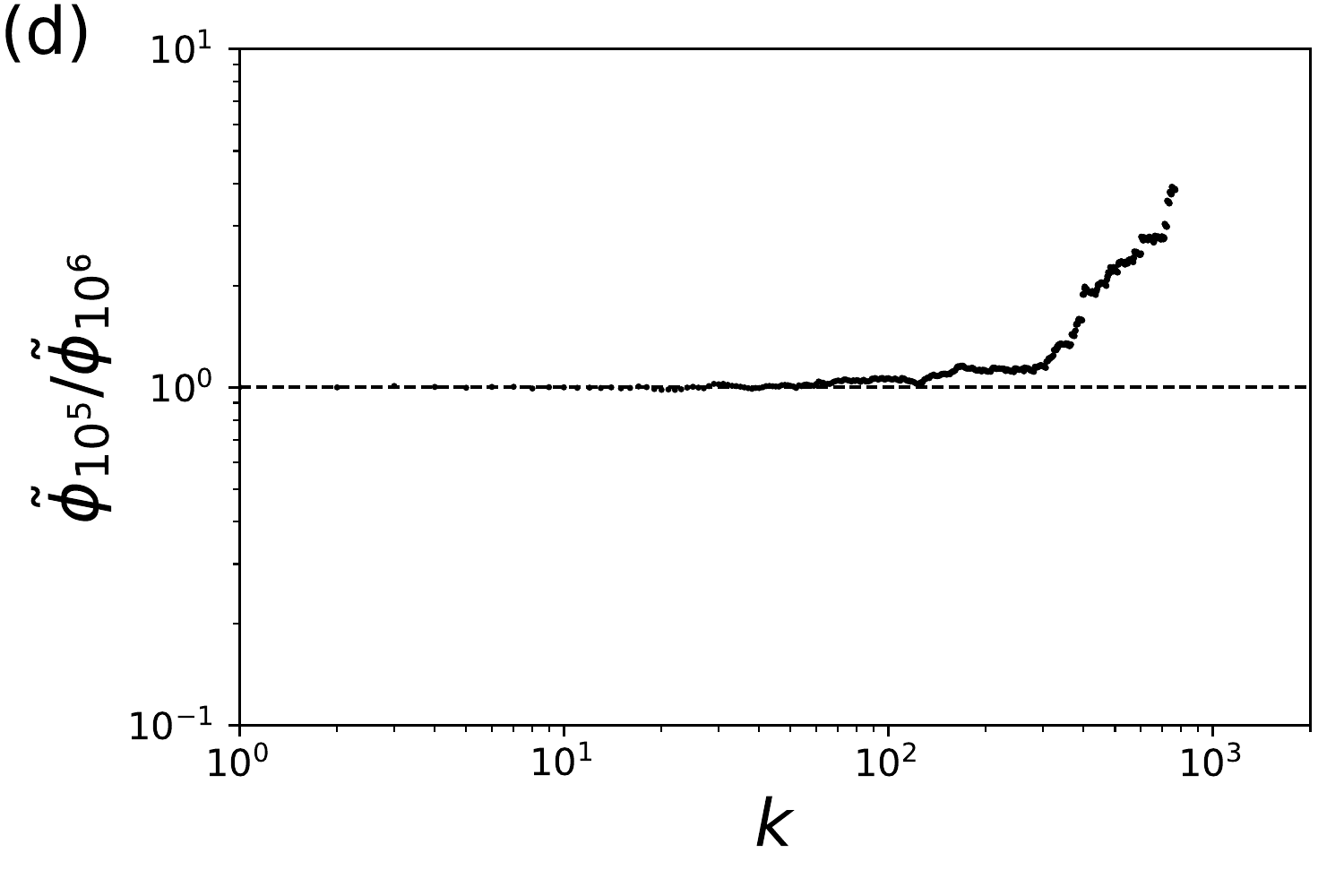}}
    \caption{An illustration that the relative attachment kernel in the BA model is \textit{time independent}. (a) The relative attachment kernel, $\phi_{t}(k,1)$, for the BA model with $m=1$ as calculated using the corrected Newman method. (b) The ratio of the calculated relative attachment kernels. (c) The cumulative sum and (d) corresponding ratio of the relative attachment kernels shown in (a). The dashed lines indicate the prediction for the BA model in (a) and (c), and the ratio expected for any time independent relative attachment kernel, which is a function of node degree only, in (b) and (d). Finite-size effects result in small deviations from the expected scaling at large $k>100$, where the data is noisy.}
    \label{fig:ba_phi_calc}
\end{figure*}
In the following, we will focus on analyzing the attachment kernel and the degree distribution of the k2 model, using the BA model as a comparison.
Each simulation is initialized with a complete graph of $m_0 = m+1$ nodes.

Figure \ref{fig:k2_degree} shows the degree distribution and the true relative attachment kernel for the k2 model with $m = 1$. Both sub-figures are averaged over 100 simulations. Early in the network development, there are only small differences between the behavior of the k2 and BA models. However, as the network grows, significant differences emerge. The duration of the initial BA-like transient phase is longer and follows the BA model even more closely for $m>1$, see \appref{ap:sims}.

Over short timescales, the network growth appears largely indistinguishable from linear preferential attachment. However, over longer timescales, the attachment kernel shows clear deviations from this simple scaling, with a plateau region at moderate degree preceding a super-linear tail. The anomalous scaling observed is most clearly seen for nodes with moderate degree  in the range $k \approx 10$ to $k \approx 300$. This region suggests that there may be multiple timescales of interest at play in the evolution of the k2 model. 

As noted, the BA model incorporates a rich-get-richer mechanism, but does not account for any mechanisms of mutual benefit between nodes; new nodes added to the network receive no benefit from attaching to a hub node as opposed to any other less important node. Conversely, in the k2 model, when a node $i$, added to the network at time $t_i$, attaches to a hub node, the new node's initial attractiveness is given by
\begin{equation}
    \Pi_i(k_i;t_i) \propto k_i^{(2)}(t_i) = 1 + k_\mathrm{hub}^{(1)}(t_i)\approx k_{hub}^{(1)}(t_i),
\end{equation}
which is completely determined by the first degree of the targeted hub. This has the counter-intuitive effect that the tail of the attachment kernel appears to show super-linear preferential attachment, implying gelation, but that the change in the number of new edges in the influence network is dominated by new nodes with degree $k^{(1)}=1$.
As a consequence, the k2 model appears to show a gelation-like phenomenon to communities rather than hubs, resulting in the plateaus shown in Fig.~\ref{fig:k2_degree}.

The true relative attachment kernel in Fig.~\ref{fig:k2_degree} is typically not accessible for a real network. To illustrate the risks this may pose, let us assume the k2 model is a real network and fit the relative attachment kernel, on the assumption that $\phi_{t}(k,1) \propto k^{\gamma}$ for a positive exponent $\gamma$. From Fig.~\ref{fig:k2_degree}(b) we may deduce that for $t=10^3$, the k2 model has an approximately linear (or possibly slightly sub-linear) attachment kernel, whereas at $t=10^6$, the attachment kernel is highly non-linear, but clearly grows faster with $k$ than the simple prediction from linear preferential attachment. If we were to use these results to infer the future scaling of the network, the data at $t=10^3$ would suggest the network might approach a stretched exponential degree distribution, whereas from the data for $t=10^6$, we might paradoxically infer the network is approaching a gelation state. In the case of the k2 model this approach is misleading, but for other networks this approach may be a good first approximation. However, what is clear is that simply calculating the attachment kernel of a network at one point in time is not sufficient to determine the form of the attachment kernel in the past or future. Likewise, since the degree distribution is determined by the underlying dynamical process growing the network, we cannot accurately know how the degree distribution will evolve in the future.

In the case of a real network, we can only estimate the relative attachment kernel by observing the degree of nodes to which new nodes added to the network attach. To simulate this real-network scenario, we apply the corrected Newman method to a single simulation of the k2 model as shown in Fig.~\ref{fig:k2_phi_calc}. Figure \ref{fig:k2_phi_calc}(a) shows the calculated relative attachment kernel for the k2 model at times $t=10^5$ and $t=10^6$. As in Fig.~\ref{fig:k2_degree}(b), nodes with moderate degree, $k \approx 30$, show an excess in the relative attachment kernel. In Fig.~\ref{fig:k2_phi_calc}(b), deviations in the relative attachment kernel are shown explicitly by taking the ratio to the relative attachment kernels at $t=10^5$ and $t=10^6$. For very small degree nodes, the ratio is approximately one indicating that the attachment kernel is time independent at these degrees. Above $k=10$, the ratio clearly deviates from one, indicating that the relative attachment kernel is time dependent. For visual clarity, Fig.~\ref{fig:k2_phi_calc}(c) \& (d) show the equivalent as (a) \& (b) but for the cumulative sum of the relative attachment kernel, defined as
\begin{equation}
    \tilde{\phi}(k,1) = \sum_{\tilde{k} \leq k}\phi(\tilde{k},1).
\end{equation}

It is important to note that the estimated attachment kernel using Newman's method is not fully consistent with the true attachment kernel; the magnitude of the excess in the attachment probabilities is much smaller using Newman's method than the true excess shown in Fig.~\ref{fig:k2_degree}(b). This is because Newman's method constructs the attachment kernel by collating multiple histograms from different times in the network evolution. The consequence is that the estimated form of the attachment kernel at $t=10^6$ is more consistent with the true attachment kernel earlier in the evolution of the k2 model, rather than the current value of the attachment kernel.

To verify that the deviations in the relative attachment kernel are due to the evolution of the k2 model and not numerical errors, we repeat the analysis shown in Fig.~\ref{fig:k2_phi_calc} for the BA model where the relative attachment kernel is expected to be time independent. Figure~\ref{fig:ba_phi_calc}(a) shows that the relative attachment kernels are effectively indistinguishable at different times in the network evolution. This is confirmed by Fig.~\ref{fig:ba_phi_calc}(b) where the ratio of the relative attachment kernels is approximately one for all nodes with degree $k < 100$. Noise in the tail obscures the ratio for $k>100$.

Overall, Fig.~\ref{fig:ba_phi_calc} indicates that using the corrected Newman method is effective, to an extent, at estimating the relative attachment kernel of a network and testing whether it exhibits time dependence. This suggests that the deviations in the relative attachment kernel observed in Fig.~\ref{fig:k2_phi_calc} are due to the structural properties of the k2 model, and not due to limitations in the method used to estimate the relative attachment kernel. Hence, we can deduce that the relative attachment kernel for the k2 model is \textit{time dependent}.

\subsection{Mathematical Results \label{sec:maths}}
Given the complexity of the k2 model, exact analytical results are hard to derive. However, using simple arguments, we can demonstrate the inconsistencies that arise from assuming the k2 model follows a simple form of non-linear preferential attachment.

From the definition of the k2 model, we can make a continuum approximation and write the evolution of the degree, $k_i^{(1)}(t)$, of a given node $i$ as
\begin{subequations}
\begin{align}
  \frac{d k_{i}^{(1)}(\mathrm{t})}{d t}
  &=
  m\Pi_{i}^{(k2)}(t),
  \label{eq:k1_evolution}
  \\
  \Pi_{i}^{(k2)}(t)
  &\equiv
  \frac{k_{i}^{(2)}(t)}{\sum_{j=1}^{N} k_{j}^{(2)}(t)} \approx \frac{k_{i}^{(2)}(t)}{\sum_{j=1}^{N} (k_{j}^{(1)}(t))^{2}},
  \label{eq:pi_denominator}
\end{align}
\end{subequations}
where node $i$ is added to the network at time $t_{i} (\leq t)$. The second degree, $k_{i}^{(2)}(t)$, is defined according to Eq.~\eqref{eq:k2def}, the summation is over all nodes in the network, and the approximation is an equality if $m=1$, see \appref{ap:maths}. We can write the evolution of the second degree as
\begin{equation}
\frac{d k_{i}^{(2)}(\mathrm{t})}{d t}\approx m\frac{\left(k_{i}^{(1)}(t)\right)^{2}+\xi_{i}(t)}{\sum_{j=1}^{N}\left(k_{j}^{(1)}(t)\right)^{2}},
\label{eq:k2_evolution}
\end{equation}
where
\begin{equation}
\xi_{i}(t) \equiv \sum_{\beta=1}^{k_{i_{1}}^{(1)}(t)-1} k_{i_{1}(\beta)}^{(1)}(t)+\cdots+\sum_{\beta=1}^{k_{i \alpha}^{(1)}(t)-1} k_{i_{\alpha(\beta)}}^{(1)}(t),
\label{eq:outside_corrs}
\end{equation}
see \appref{ap:maths} for a derivation. Here $i_{\alpha(\beta)}$ represents the node $\beta$ connected to node $i_{\alpha}$. Equation~\eqref{eq:outside_corrs} represents the effect of non-neighboring nodes on node $i$. Since the first degree of a node can only grow over time, $\xi_{i}(t)$ is a positive semi-definite monotone increasing function with respect to time $t$, that is,
\begin{subequations}
\begin{align}
\xi_{i}(t)
& \geq 0,
\label{eq:xi_condition1}
\\
\frac{d}{d t} \xi_{i}(t) & \geq 0.
\label{eq:xi_condition2}
\end{align}
\end{subequations}

Our aim in the following is to write $\xi_{i}(t)$ as a function of the first degree, $k^{(1)}$, only. To do so we rearrange Eq.~\eqref{eq:k2_evolution} to make $\xi_{i}(t)$ the subject and substitute in Eq.~\eqref{eq:k1_evolution} and Eq.~\eqref{eq:pi_denominator},
\begin{multline}
\xi_{i}(t)=\frac{1}{m}\left(\sum_{j=1}^{N}\left(k_{j}^{(1)}(t)\right)^{2}\right)\cdot\\\frac{d}{dt}\left[\frac{1}{m}\left(\sum_{j=1}^{N}\left(k_{j}^{(1)}(t)\right)^2\right)\cdot \frac{d k_{i}^{(1)}(\mathrm{t})}{d t} \right]-\left(k_{i}^{(1)}(t)\right)^{2},
\label{eq:xi_rearrange2}
\end{multline}
which is a function of the first degree only. Here we note that the summations in Eq.~\eqref{eq:xi_rearrange2} correspond to the denominator in Eq.~\eqref{eq:pi_denominator}. This is the sum over $k_{i}^{(2)}(t)$ for each node in the network, and hence, corresponds to the twice the total number of edges in the influence network, which we label as
\begin{equation}
    E^{(2)}(t) = \frac{1}{2}\sum_{j=1}^{N(t)}
 k^{(2)}_j(t)
 \approx
 \frac{1}{2}\sum_{j=1}^{N(t)}
 (k^{(1)}_j(t))^2.
 \label{eq:influence_edges}
\end{equation}

Figure~\ref{fig:k2_degree} shows the degree distribution and the relative attachment kernel for the k2 model obtained from simulations. As a thought experiment, let us suppose that these simulations are not for a theoretical network model but that the data represents a real world network. For the network at small times in its evolution, the degree distribution and the attachment kernel are closely approximated by the BA model.

For preferential attachment (linear or non-linear), it is known that, on average, the degree of a given node $i$ evolves in time as a power function given by
\begin{equation}
k_{i}^{(1)}(t)=m\left(\frac{t}{t_{i}}\right)^{\delta},\ t\geq t_i,
\label{eq:power_function}
\end{equation}
where $t_i$ is the time at which node $i$ was added to the network, and $\delta = 1/2$ for linear preferential attachment \cite{newman2018}. In the case of sub- (super-) linear preferential attachment, $\delta < 1/2$ ($\delta > 1/2$). Let us assume Eq.~\eqref{eq:power_function} holds and test whether this simple scaling is consistent with the mathematical form of the k2 model. First, we substitute Eq.~\eqref{eq:influence_edges} into Eq.~\eqref{eq:xi_rearrange2},
\begin{multline}
 \xi_i(t)=
 \frac{4}{m^2}
 E^{(2)}(t)
 \frac{d}{dt} \left[
 E^{(2)}(t)
 \frac{dk^{(1)}_i(t)}{dt}
 \right]
 - (k^{(1)}_i(t) )^2.
\label{eq:xi_subbed1}
\end{multline}
In the case of the k2 model where one node is added to the network at each time step, we initialize our network such that $t_j = j$ and note that the number of nodes in the network at time $t$ is given by $N(t)=m_0+t\approx t$ for large $t$.

Using this initialization, we now calculate the value of $E^{(2)}(t)$ by approximating the sum as an integral and substituting in Eq.~\eqref{eq:power_function},
\begin{equation}
\begin{aligned}
        E^{(2)}(t)
 &\approx
 \frac{1}{2}\sum_{j=1}^{t}
 (k^{(1)}_j(t))^2
 \\
 &\approx
 \frac{m^2}{2} \int_{1}^{t} dt' \;
 \left(\frac{t}{t'}\right)^{2\delta}
 \\
 &\approx
 \frac{m^2 t^{2\delta} }{2(1-2\delta) }
 \left[  (t')^{1-2\delta}  \right]_{1}^{t}.
 \label{e:E2def}
\end{aligned}
\end{equation}

There are three cases for the different possible values of $\delta$: Case (i) $2\delta<1$.  Corresponding to sub-linear preferential attachment, this scenario is likely to be irrelevant for the k2 model since the influence network cannot grow slower than the original network in the BA model. In this case we expect to find linear growth in the number of edges in the influence network
\begin{equation}
 E^{(2)}(t)
 \approx
 \frac{m^2}{2(1-2\delta)} t.
 \label{e:E2i}
\end{equation}
Here $E^{(2)}(t)$ is dominated by the youngest nodes (created at the largest times $t_i$) as the older nodes grow too slowly.

Case (ii) $2\delta=1$. Corresponding to linear preferential attachment, this is the case for the BA model,
\begin{equation}
E^{(2)}(t)
 \approx
 \frac{m^2}{2} \, t \ln(t).
 \label{e:E2ii}
\end{equation}

Case (iii) $2\delta>1$.  Corresponding to super-linear preferential attachment where there is some enhancement over linear preferential attachment. For the k2 model, this scenario is plausible since we know that for any given node $k^{(2)}_i(t) \geq k^{(1)}_i(t)$. In this case we find
\begin{equation}
    E^{(2)}(t)
 \approx
 \frac{m^2}{2(2\delta-1)} (t^{2\delta} - t),
 \label{e:E2iii}
\end{equation}
where the growth in the number of edges in the influence network is dominated by the oldest nodes in the network.

Let us assume case (iii) is valid for the k2 model. Substituting Eq.~\eqref{e:E2iii} into Eq.~\eqref{eq:xi_subbed1} we find,
\begin{equation}
    \begin{aligned}
        \xi_i(t)
 &=
 m^3
 \frac{\delta}{(2\delta-1)^2}
 \left(\frac{1}{t_i}\right)^\delta \cdot
 \\
 &\; \;
 \left[
 (3\delta-1)
  t^{5\delta-2}
  -
  (4 \delta - 1) t^{3\delta - 1}
  +
  \delta t^\delta
  \right]
 - m^2 \left(\frac{t}{t_i}\right)^{2\delta}
 \\
 &=
 a_1
  t^{5\delta-2}
  +
  a_2 t^\delta
  -
  a_3 t^{3\delta - 1}
  -
  a_4 t^{2\delta},
  \label{eq:four_terms}
    \end{aligned}
\end{equation}
where in the final line we have grouped all the constants for each term into a single positive prefactor, $a_{1}$ to $a_4$.

Recall that the k2 model requires that $\xi_i(t)$ is a positive, semi-defined monotonically increasing function, and note that Eq.~\eqref{eq:four_terms} is only valid for $\delta > 1/2$. As $t\rightarrow \infty$, the first term of Eq.~\eqref{eq:four_terms} will dominate the second if $5\delta - 2 \geq \delta$, $\delta \geq 1/2$. Likewise, the third term will dominate the fourth if $3\delta -1 \geq 2\delta$, $\delta \geq 1$. Hence, as $t\rightarrow \infty$, the first term is the dominant positive term and the fourth term is the dominant negative term. To ensure $\xi_i(t) \geq 0$ for all $t > t_i$, this requires the first term to grow faster than the fourth term giving $5\delta -2 \geq 2\delta$, corresponding to $\delta \geq 2/3$.

Returning to Eq.~\eqref{eq:k2_evolution} and substituting in Eq.~\eqref{eq:power_function} and Eq.~\eqref{e:E2iii}, we can also write
\begin{equation}
\frac{d k_{i}^{(2)}(\mathrm{t})}{d t}=m(2\delta -1)\frac{m^2\left(t/t_i\right)^{2\delta}+\xi_{i}(t)}{m^2(t^{2\delta} - t)}.
\label{eq:k2_evolution_subbed}
\end{equation}
We have established that to satisfy Eq.~\eqref{eq:xi_condition1}, the leading term of $\xi_i(t)$ must scale as $t^{5\delta - 2}$, and $\delta \geq 2/3$. However, as a consequence of the rules of the k2 model, at time $t > t_i$, node $i$ can gain no more than $m$ new edges in the influence network in any given time step (i.e., $k_i^{(2)}(t+1) - k_i^{(2)}(t) \leq m$). Hence, strictly for $t > t_i$, we require
\begin{equation}
\frac{d k_{i}^{(2)}(\mathrm{t})}{d t}\leq m,
\label{eq:k2_evolution_inequality}
\end{equation}
which is only satisfied if the denominator of Eq.~\eqref{eq:k2_evolution_subbed} grows at least as fast as the numerator of Eq.~\eqref{eq:k2_evolution_subbed}. This requires $t^{2\delta} \geq t^{5\delta -2}$ as $t \rightarrow \infty$. Hence, $2\delta \geq 5\delta -2$, giving $\delta \leq 2/3$. Combining the conditions in Eq.~\eqref{eq:xi_condition1} and Eq.~\eqref{eq:k2_evolution_inequality}, we find that a power function of the form given in Eq.~\eqref{eq:power_function} can only satisfy the requirements of the k2 model if $\delta = 2/3$.

\begin{figure}[h!]
    \centering
    \includegraphics[width=\linewidth]{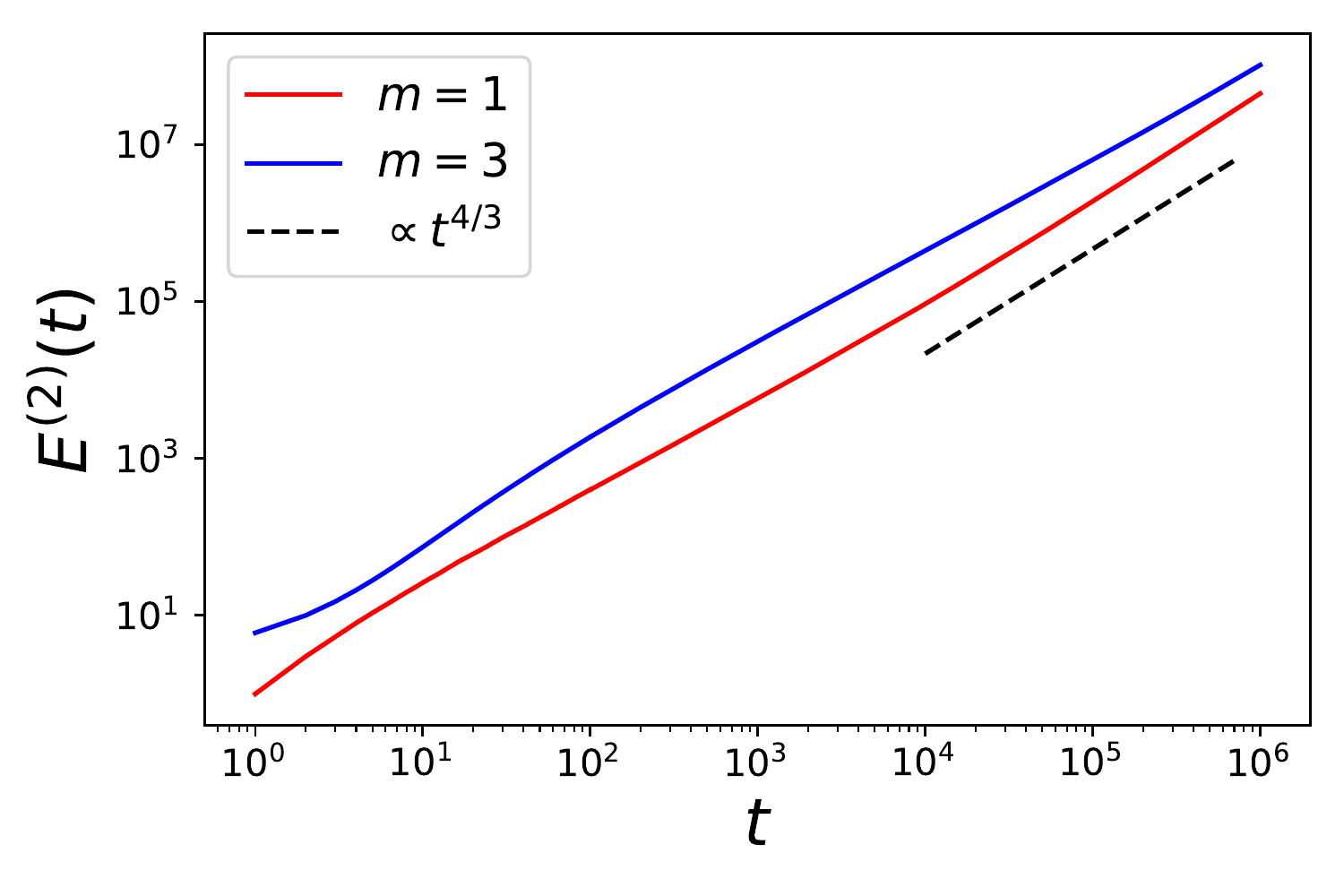}
    \caption{The number of edges in the k2 influence network. Data is averaged over 100 simulations. Error bars are negligible. For large $t$, the number of edges scales approximately with $t^{4/3}$, as indicated by the dashed line.}
    \label{fig:influence_edge_growth}
\end{figure}

To test the validity of our argument, we simulate the growth in the number of edges for the k2 influence network. This is shown in Fig.~\ref{fig:influence_edge_growth} for $m=1$ \& $3$. The figure shows that, at large $t$, the number of edges in the influence network scales as approximately $t^{4/3}$ corresponding to $\delta = 2/3$, in agreement with our prediction.

\vspace{3pt}
\vspace{3pt}

\begin{figure}[h!]
\centering
\subfloat{\includegraphics[width = \linewidth]{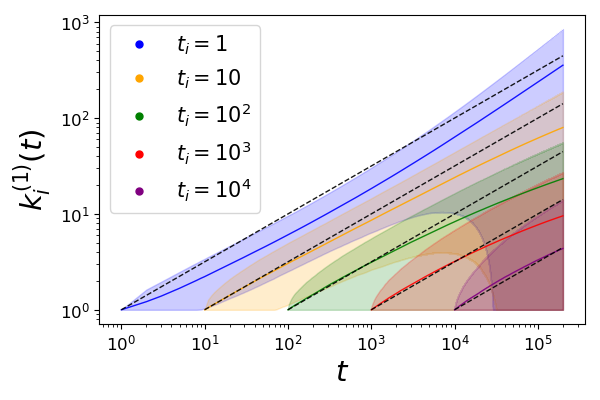}}
\caption{The evolution of the degree of individual nodes added at time $t_i$ for the k2 model with $m=1$. The shaded region around the solid lines indicates the standard deviation across 10000 simulations. The dashed lines indicate the scaling expected for the BA model, see Eq.~\eqref{eq:power_function}.}
\label{fig:degree_evo}
\end{figure}

However, further analysis appears to contradict this conclusion. Firstly, we can simulate the k2 model and track the degree of specific nodes over time, see Fig.~\ref{fig:degree_evo}. The data has been averaged over $10^4$ simulations with the shaded regions indicating the standard deviation; only with a very large sample size can the average evolution of node $i$ be observed. In most simulations, a node hardly grows at all, whereas in a few simulations, nodes grow very quickly.

\vspace{3pt}
\vspace{3pt}

For a transient period after being added to the network, the average degree evolution of a node appears to scale as $t^{1/2}$ which is the expected scaling for linear preferential attachment. This appears to contradict the $\delta = 2/3$ scaling identified previously, although we note that the integral in Eq.~\eqref{e:E2def} is dominated by the oldest nodes in the network for $\delta > 1/2$, which do appear to grow faster than $t^{1/2}$ towards the end of the simulation.

For newer nodes, after a transient period, the degree evolution appears to deviate from $t^{1/2}$ scaling, but the scaling appears to transition to $\delta <1/2$ rather than $\delta>1/2$. The time over which this transition takes place increases with the time nodes are added to the network.

This suggests that the true functional form for the degree evolution in the k2 model involves two competing terms, the first scaling as $t^{2/3}$ which is suppressed by $t_i$, and a second term which scales as $t^{1/2}$ which is suppressed by $t$.
We hypothesize, but at this stage cannot prove, that this implies two scaling regimes: For fixed $t_i$ and $t\rightarrow \infty$, the scaling of the degree evolution is dominated by a $t^{2/3}$ term to ensure that $E(t) \propto t^{4/3}$ as $t \rightarrow \infty$. For $t_i \rightarrow \infty$ and $t = t_i + \epsilon$ where $\epsilon \ll t_i$, the degree evolution of node $i$ is dominated by a $t^{1/2}$ term. Competing regimes of this type are not seen in standard non-linear preferential attachment.

It is interesting to consider the origin of the $t^{1/2}$ scaling. Our results are inconclusive, however, if we let $t_i \rightarrow \infty$ and set $t = t_i + \epsilon$ with $\epsilon \ll t_i$, a Taylor expansion of Eq.~\eqref{eq:four_terms} gives
\begin{equation}
\xi_i(t_i)
 =
b_1 t_i^{4\delta - 2}
  -
b_3 t_i^{2\delta - 1}
+
\mathcal{O}(\epsilon),
\label{eq:taylor}
\end{equation}
with positive constants $b_1$ and $b_3$, revealing that $\delta \geq 1/2$, rather than $\delta \geq 2/3$, is sufficient for ensuring that $\xi_i(t=t_i) \geq 0$ as $t_i \rightarrow \infty$.

The mathematical argument presented here does not prove the limiting behavior of the k2 model. However, the result does indicate that the inclusion of simple nearest neighbor correlations in network growth can effect the scaling of key observables. Despite initially appearing to grow as linear preferential attachment, this simple scaling breaks down as the network grows. In the case of real networks this may happen at an early stage in the evolution of a network. However, as illustrated by the k2 model, the transient time during which the model appears to grow according to linear preferential attachment may be significant - it is not uncommon to analyze real networks with $10^4-10^5$ nodes, yet in the case of the k2 model, particularly for $m>1$, see \appref{ap:sims}, the network is still in this transient period. In cases like the k2 model with complex growth rules, oversimplified assumptions derived from non-linear preferential attachment do not reflect reality.

\subsection{Application to real world networks}
\begin{figure*}
    \centering
    \subfloat{\includegraphics[width =0.5 \linewidth]{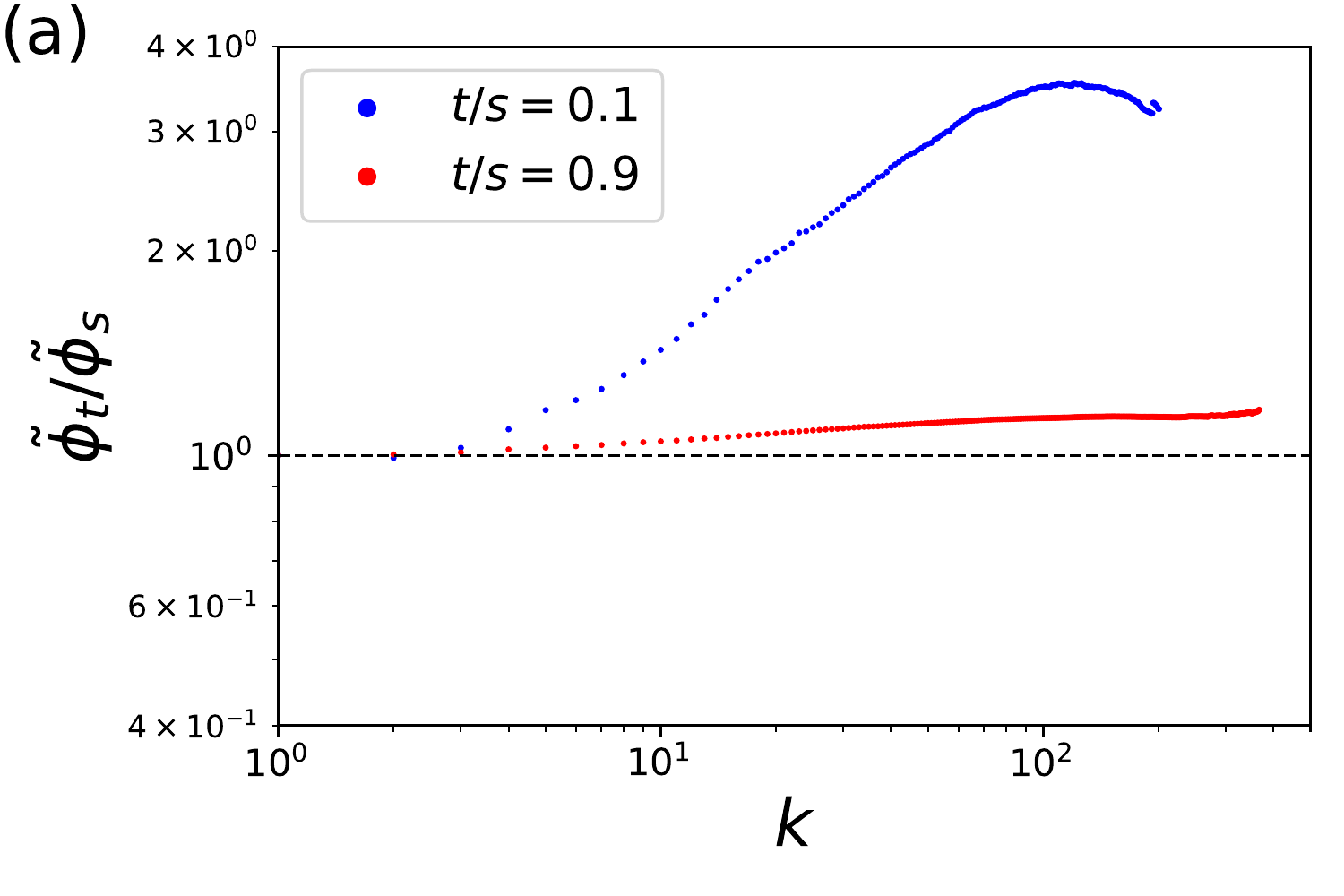}}
    \subfloat{\includegraphics[width = 0.5 \linewidth]{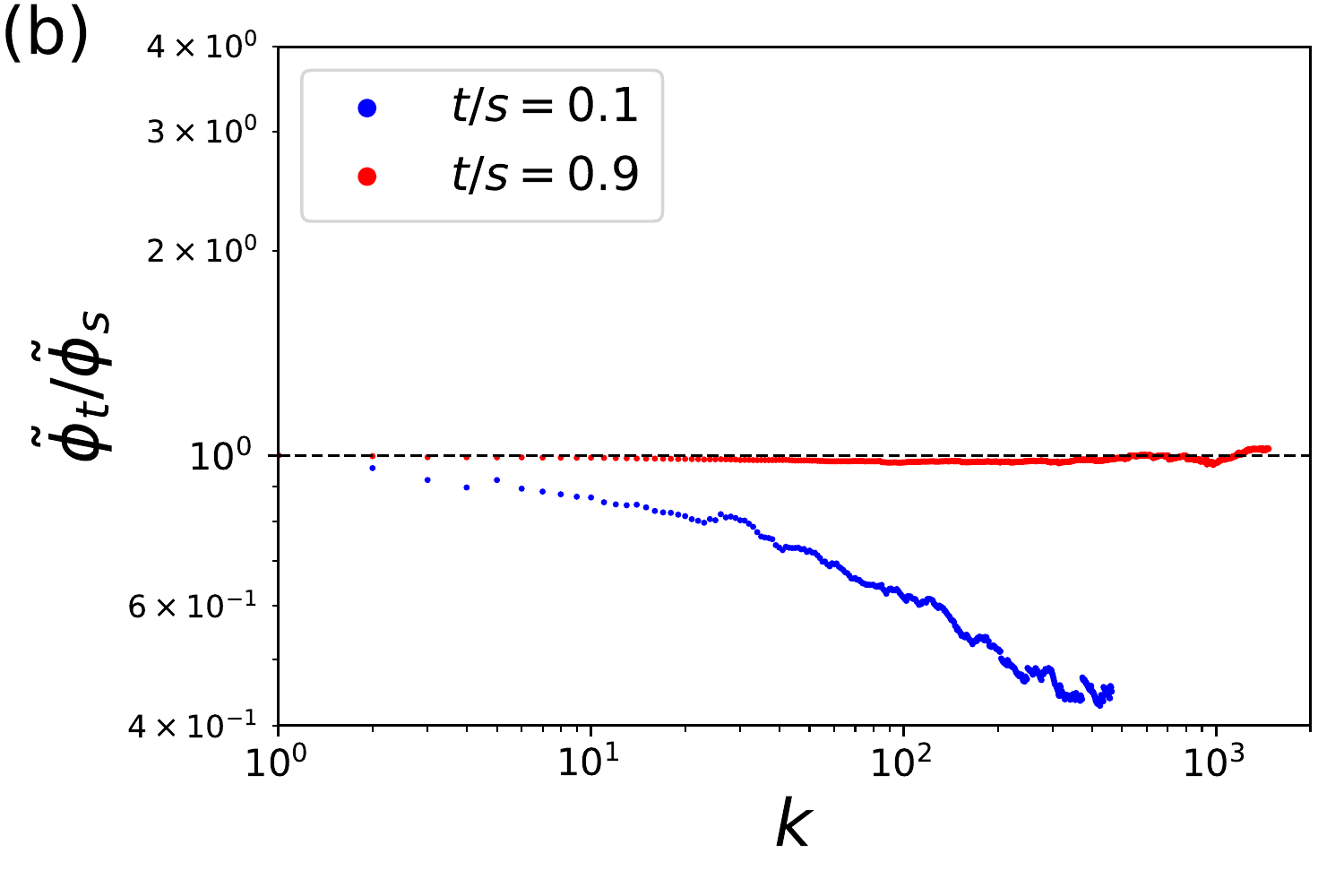}}
    \newline
    \subfloat{\includegraphics[width = 0.5 \linewidth]{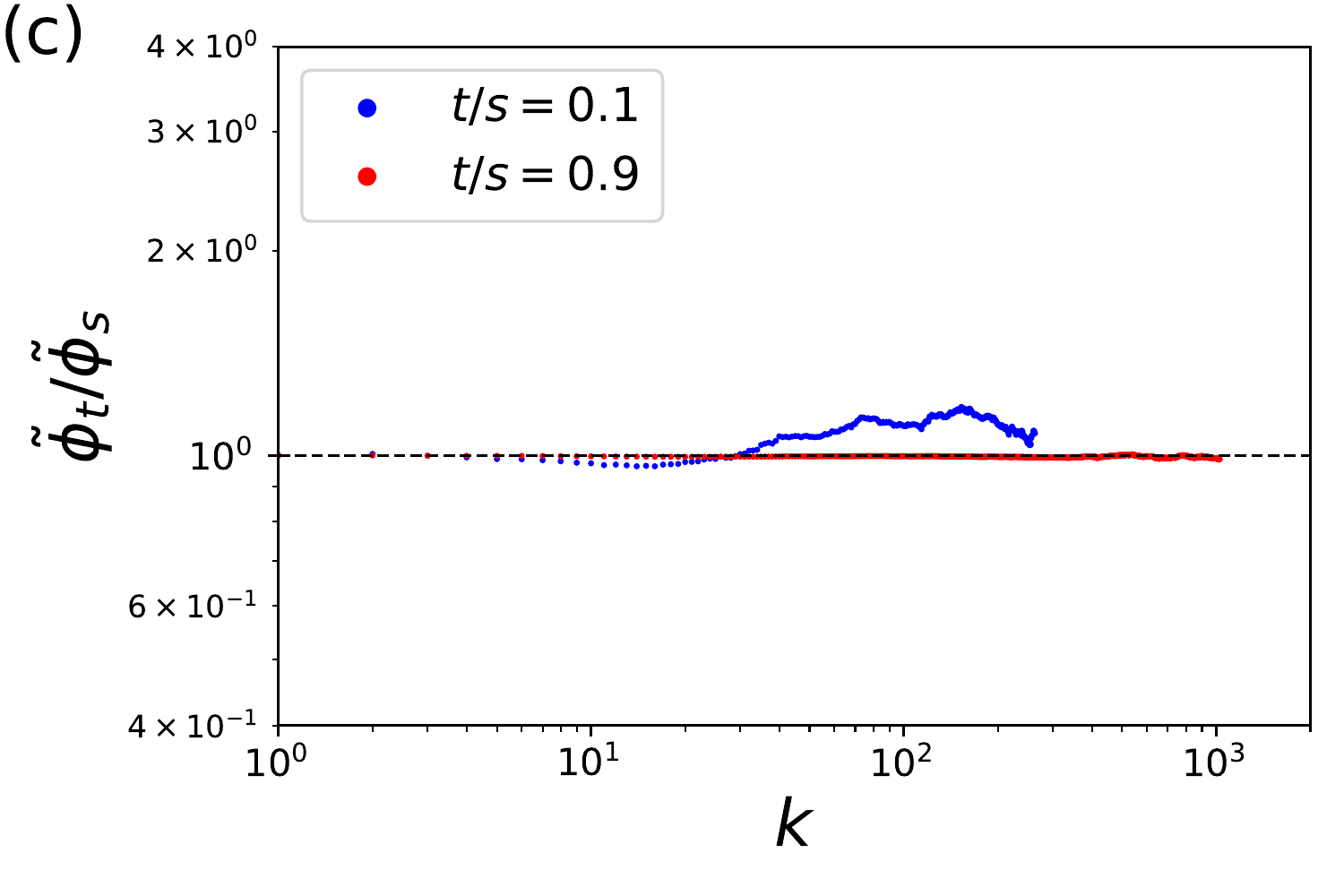}}
    \subfloat{\includegraphics[width = 0.5 \linewidth]{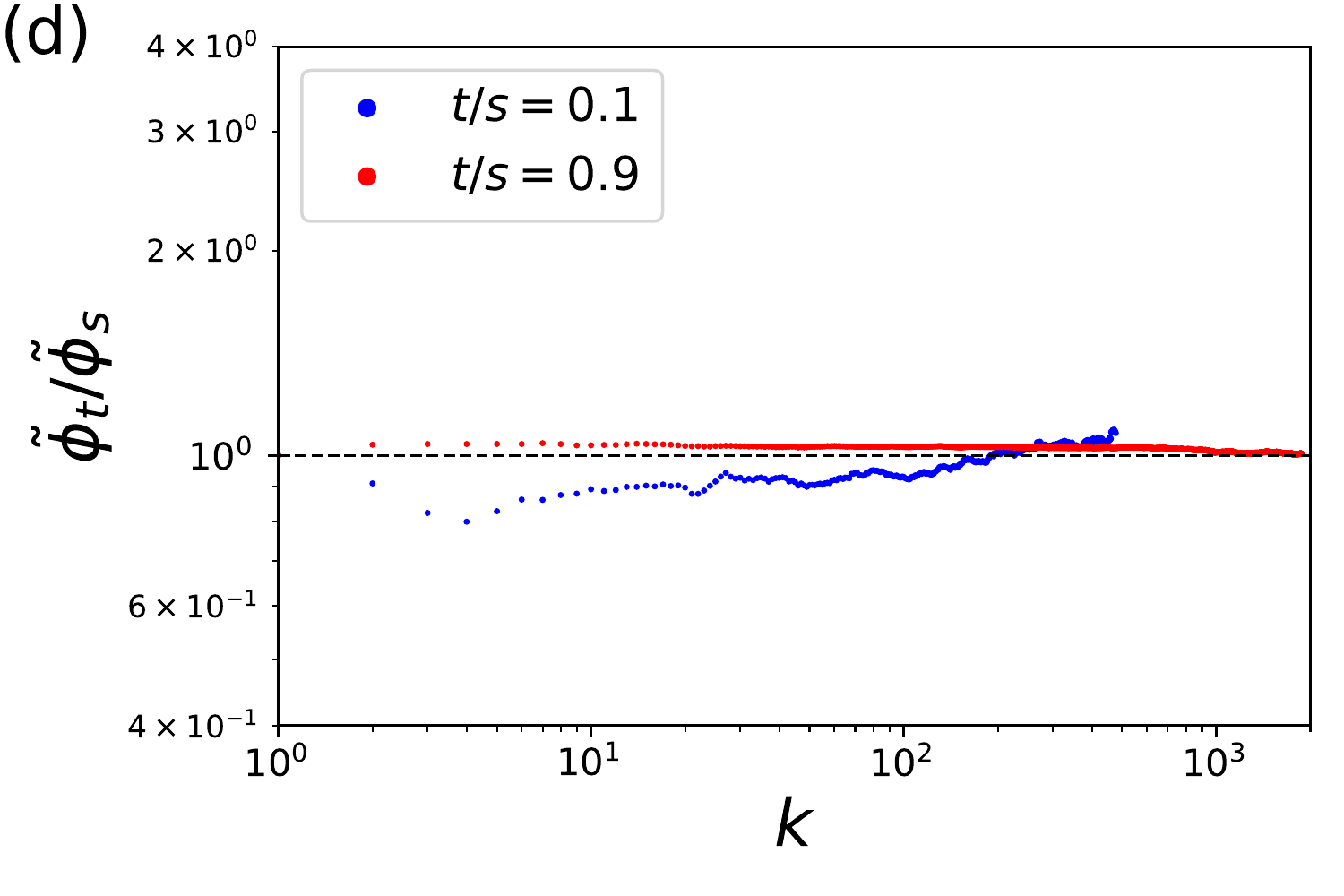}}
    \newline
    \subfloat{\includegraphics[width = 0.5 \linewidth]{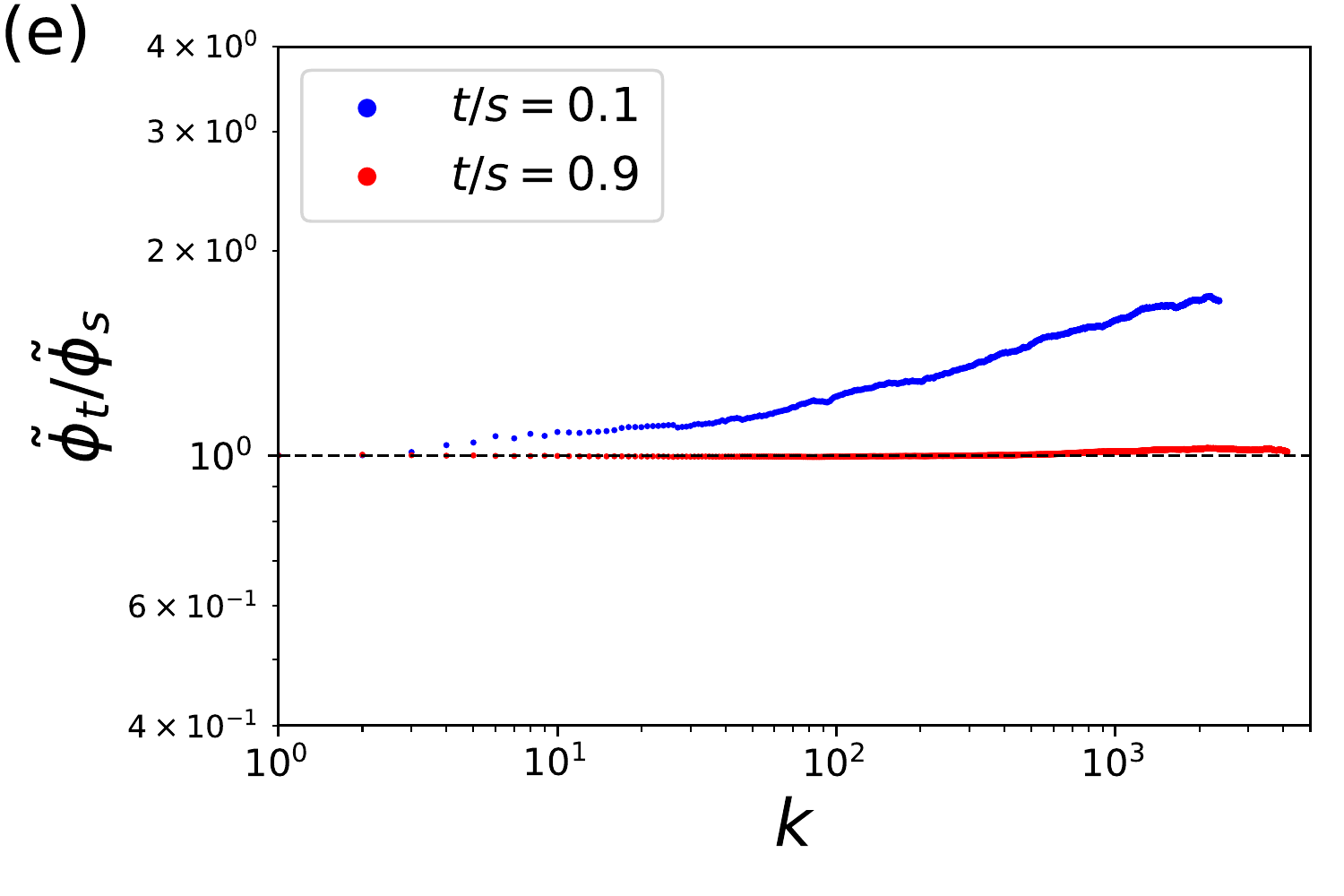}}
    \subfloat{\includegraphics[width = 0.5 \linewidth]{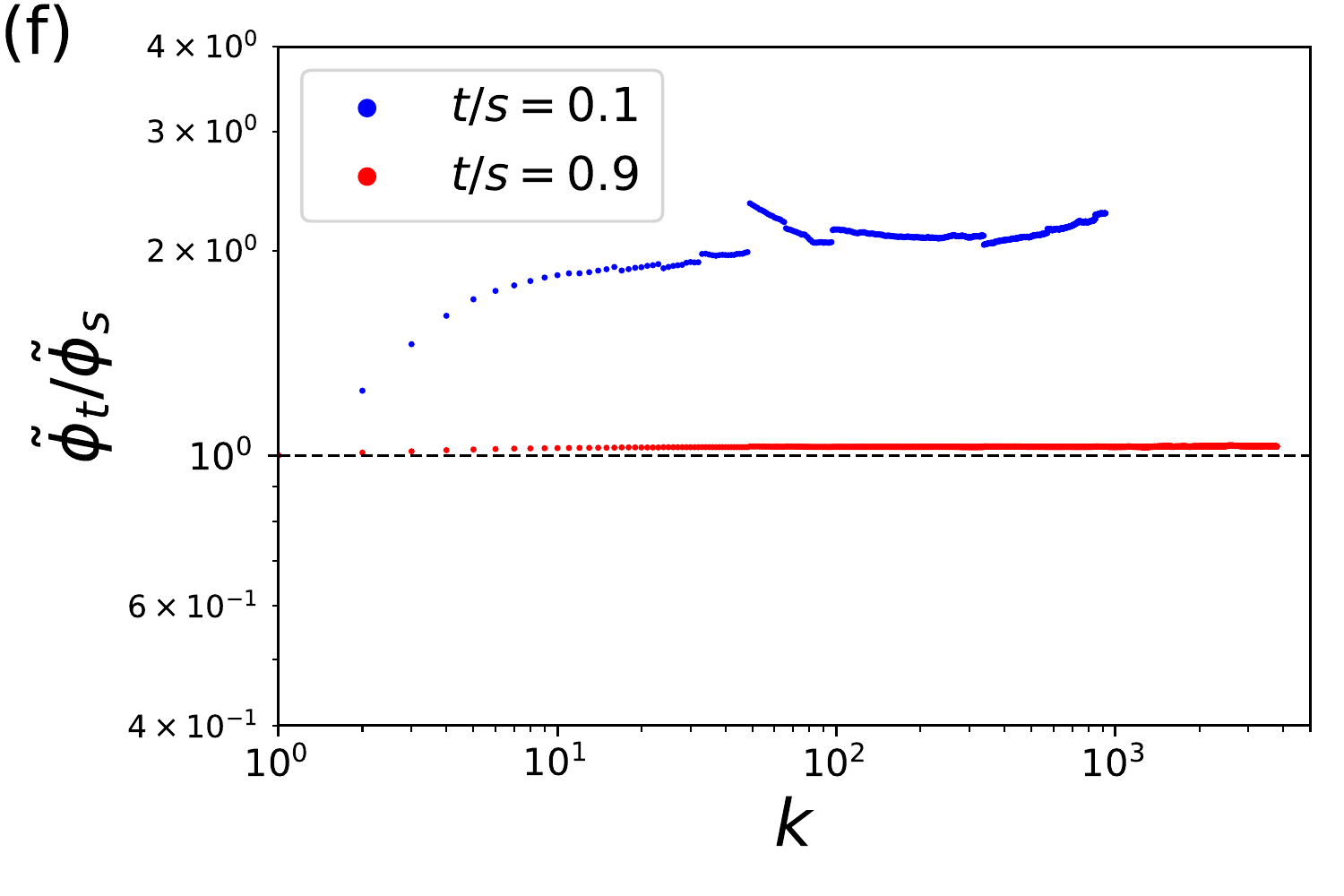}}
    \caption{The ratio of the cumulative relative attachment kernels for six real world networks; (a) Facebook friendships \cite{viswanath2009}, (b) Youtube followers \cite{mislove2009}, the APS citation network \cite{aps}, (d) \texttt{hep-ph} arXiv collaborations \cite{konect2013}, (e) the Flickr network \cite{mislove2009}, and (f) hyperlinks on English Wikipedia \cite{mislove2009}. For all networks the ratio is shown for early, $t/s = 0.1$, and late, $t/s = 0.9$, in the evolution of the recorded network relative to the endpoint, $s$; time is measured in the net number of edges added to the network. }
    \label{fig:real_networks}
\end{figure*}
Figure~\ref{fig:real_networks} shows the ratio of the cumulative relative attachment kernels for six real world growing networks: (a) a regional friendship network on Facebook \cite{viswanath2009}, (b) the Youtube follower network \cite{mislove2009}, (c) the APS citation network \cite{aps}, (d) the \texttt{hep-ph} arXiv collaboration network \cite{kunegis2013}, (e) the Flickr follower network \cite{mislove2009}, and (f) the hyperlink network for English Wikipedia \cite{mislove2009}. For all networks, the cumulative relative attachment kernel is calculated at two time points early and late in the network's evolution relative to the endpoint of the dataset. In some cases the full evolution of the network is not known and is accounted for with a large initial graph at $t=0$.

Two details are clear in Fig.~\ref{fig:real_networks}. Firstly, on short timescales the relative attachment kernels are approximately time independent, with only small deviations observed. However, over longer time periods, the ratio  of the relative attachment kernels is not constant indicating time dependence.

There is significant diversity in the changes observed to the ratio  of the relative attachment kernels. In Fig.~\ref{fig:real_networks}(a), (e) and (f), the ratio is (to a good approximation) monotonically increasing. This implies that if we were to approximate the attachment kernel of these networks using non-linear preferential attachment, $\Pi(k)\propto k^{\gamma}$, the exponent $\gamma$ will appear to have reduced over time (the network growth is becoming more sub-linear).
Conversely, in Fig.~\ref{fig:real_networks}(b) we see the opposite effect where the ratio is approximately monotonically decreasing, implying an increase in the exponent $\gamma$ (the network growth is becoming more super-linear). In the cases of Fig.~\ref{fig:real_networks}(a), (b) and (e), the form of non-linear attachment (i.e.,  sub-linear, $\gamma < 1$, or super-linear, $\gamma > 1$) does not change. However, in the case of the Wikipedia hyperlink network in Fig.~\ref{fig:real_networks}(f), the relative attachment kernel early in the network's evolution appeared super-linear, whereas by its endpoint, the attachment kernel was measured to be sub-linear. Here we reiterate our previous criticism: if the attachment kernel of a network is meant to predict its future evolution, how do we reconcile that measurements across some time windows result in one prediction, while other time windows result in a different, wholly incompatible prediction.

The story in Fig.~\ref{fig:real_networks}(c) and (d) is more complex. In both cases, the ratios of the relative attachment kernels initially appear to decrease below $1$, before increasing at moderate degree and exceeding a ratio of $1$ at large degree. This behavior is qualitatively very similar to the dynamics observed in Fig.~\ref{fig:k2_phi_calc} for the k2 model. In these cases, approximating the change in the attachment kernel as a change in the non-linear preferential attachment kernel $\gamma$ is not easy since the data for small degree may imply an increase in $\gamma$ whereas the data for large degrees may imply a decrease in the exponent $\gamma$. In such cases, simple assumptions of non-linear preferential attachment are not sufficient to draw reliable conclusions about the future evolution or past origin of a network.

A number of mechanisms may be responsible for the appearance of time dependence in the relative attachment kernel. Generally, the simplest explanations for time dependence in network growth relate to changes in network parameters over time. In most network growth models (including in the k2 model) these are assumed to be time independent for simplicity. An example of such a parameter includes the number of edges added by each new node added to a network, $m$.

In the BA model, the limiting degree distribution is given by
\begin{equation}
    p_{\infty}(k) = \frac{2m(m+1)}{k(k+1)(k+2)},
\end{equation}
where we note that this solution is valid for sufficiently large graphs given any initial network at time $t_0$. As a result, if we let $m \rightarrow m(t)$, it is plausible that we will observe transient behavior during which the form of the degree distribution may change over time. The same argument holds for measurements of the network attachment kernel. Such time dependence in $m$ is not hypothetical and has been shown to be true by \citeauthor{leskovec2005} \cite{leskovec2005} for a number of different growing networks including citation networks, patent networks and affiliation networks. Some network models consider growth in the average out-degree of nodes over time, see for instance \cite{barabasi2002}. However, despite showing complex time dependent scaling in the time evolution of individual nodes (the analytical form for the degree distribution is not solvable), the authors argue that the limiting degree evolution implies power-law scaling in the degree distribution for large graphs. This may be true, but as the k2 model illustrates, in some cases the transient phase of a network may be so long such that, for all practical purposes, the limiting degree distribution is not necessarily observed during the lifetime of a real world network. Slow convergence to a limiting degree distribution has been noted previously for node copying models in \cite{ispolatov2005}.

Another simple parameters which may effect the time dependence of either the degree distribution or attachment kernel of a network is the exponent for non-linear preferential attachment. Consider letting $\gamma \rightarrow \gamma(t)$ for $\Pi (k) \propto k^{\gamma}$. It is already known that $\gamma$ effects the limiting degree distribution of non-linear preferential attachment \cite{krapivsky2000}, and that for $\gamma > 1$, the degree distribution appears scale free for a transient period \cite{krapivsky2008}. Hence, any variation in $\gamma$ is likely to add to the complexity of the time dependence observed in network growth observables.

So far, we have provided only two examples of parameters whose time dependence may alter the transient behavior of a network's degree distribution or attachment kernel. However, we would argue that any solvable network model where a constant parameter appears in the analytical form for the degree distribution is likely to exhibit time dependent transients if that constant becomes time dependent itself.

In cases like the k2 model, where no individual parameter has been set to be time dependent, the reasons for complex scaling in the observables of network growth are less easily explained and may not be easy to elucidate from data. In the k2 model, the origin lies in the implicit time dependence of the local network structure which results in super-linear scaling in the influence network, associated with gelation to important communities. However, any network growth model where the attachment kernel is determined by an observable which implicitly changes over time as the structure of the network changes is likely to exhibit similar time dependence. For instance, network growth based on attaching to nodes according to their betweenness is likely to exhibit a time dependent network attachment kernel, as investigated by \cite{Topirceanu2018}.

We note to the reader that it was our original intention to perform the analysis in Fig.~\ref{fig:real_networks} with statistical rigor. However, this task has proven difficult given that (1) our data breaks many of the assumptions underlying common statistical tests, and (2) it is not yet fully understood how techniques for estimating attachment kernels, like Newman's method, are effected/biased by time dependence -- by construction, these techniques assume the attachment kernel is time independent. We highlight the need to tackle these problems with better statistical network analysis in future work.

\section{Discussion \& Conclusion}
The study of complex networks has come to dominate complexity science in the 21st century, and is likely to become more prominent in a hyper-connected world. Not only have complex networks become influential in physics and mathematics, but their trans-disciplinary appeal has led to their use across almost all areas of science and academia, from archaeology \cite{knappett2008} to neuroscience \cite{bassett2017}, economics \cite{schweitzer2009} to epidemics \cite{pastor2015}, and many more.

A key feature of network science is the study of how networks emerge and evolve over time, and numerous models and techniques have been developed to explore this problem \cite{barabasi1999,krapivsky2000,newman2001,vazquez2000,KR01,barabasi2002,jeong2003,V03,FLDG03,SK04,ES05,GAE15,pham2015,newman2018,barabasi2016}. In almost all cases, these models and techniques have their limitations and are only applicable to the real world under a number of key constraints. Despite this, the spread of network science has been so extensive that many of these approaches are being used without a robust understanding of their underlying assumptions. In this paper we have discussed two such assumptions: (1) that the rules underlying network growth do not depend on time, and (2) that the degree of nodes in a network is the key observable determining network evolution.

The number of research papers discussing network growth and attempting to infer their underlying mechanisms is vast, often guided by simple network models to inform their analysis. However, the models most frequently discussed in popular network science textbooks, for instance \cite{newman2018,barabasi2016}, almost always assume that underlying growth rules are fixed in time. As a result, it is not particularly surprising that most papers inferring network growth mechanisms also make these assumptions. In many contexts such assumptions are sensible and essential, allowing for analytically tractable calculations which may otherwise be impossible. However, in some real world scenarios such approaches may not be suitable. A selection of papers which do consider the implication of these assumptions include \cite{sun2020,dangalchev2004} in the context of preferential attachment models, \cite{amati2019,amano2018} in organizational networks, \cite{Topirceanu2018} in social networks, and others \cite{Wang2018,lambiotte2016,bhat2016}.

In this paper, we have tried to highlight how very simple network growth rules can break both the time-independence of the network degree distribution and the time-independence of the node-node attachment probability. We have done this by introducing the k2 model, a simple variant of the Barab{\'a}si-Albert model where the attractiveness of a node is correlated to the attractiveness of a node's neighbors. Even though such a network growth rule does not contain an explicit time dependence, the formation of clusters means that a node's attractiveness is implicitly time dependent through its dependence on its local environment. This mechanism is relevant for real-world networks involving mutual benefit where a node gains an advantage from being connected to an influential neighbor, such as in collaboration networks \cite{Li2019}, or citation networks \cite{vazquez2000}, or indirectly in systems with neighbor-neighbor interactions and copying processes \cite{ohno2013,kang2020,bhat2016,lambiotte2016,GAE15,toivonen2009,ispolatov2005,kim2002,granovetter1977}.

The k2 model shows that for small networks, the degree distribution appears approximately power-law, and the attachment kernel is approximately linear, both of which are consistent with preferential attachment. However, after a lengthy transient period, both the degree distribution and attachment kernel show significant deviations from the simple scaling predicted for preferential attachment. These deviations grow over time showing strong time dependence. We support these findings with an approximate analytical treatment showing that assumptions of simple scaling forms in the evolution of individual nodes in the k2 model results in inconsistencies in the mathematics, suggesting that numerous scaling regimes are interacting and changing over time.

The k2 model is an idealized network growth model -- it does not reflect real-world networks, even if the underlying mechanism has explanatory value. However, changes in the degree distribution and the attachment kernel can also be seen in real data of varying origins. In six networks for which dynamic network data is available (three social networks, one hyperlink network, one collaboration network, and one citation network), we have found that these networks are approximately time independent on short timescales, but show significant time dependency, and diversity in that dependency, over longer timescales. In some cases, this time dependency may have simple origins such as node aging, changes in the average out-degree over time, or changes in the exponent for preferential attachment. However, in some cases, time dependency may arise implicitly resulting in complex scaling.

In the context of the wider debate on ``scale free'' networks, it is worth considering the following. If the generative mechanisms underlying network growth are not constant in time, is it plausible that the functional form for network degree distributions will be constant in time? In many cases the change over time may be very small. However, if we apply a strict definition of ``scale freeness,'' small changes in the degree distribution may be sufficient to induce changes in the most-likely functional form for the degree distribution as predicted using the current state of the art measures \cite{broido2018}. If this is in fact the case, this may explain why only $4\%$ of real world networks have been identified as scale free \cite{broido2018}.

To conclude, as long as network science techniques are being applied to the real world by experts and non-experts alike, it is essential that we understand the limitations of simple models and consider their underlying assumptions. Here, we have shown how very simple, sociologically meaningful changes to network growth models can profoundly effect both the time dependence of network growth, and the assumption that node degree determines network evolution.

While the k2 model serves an illustrative purpose, the ideas drawn from its evolution apply to real networks, which show diverse time dependence over extended durations. While this appears to be a disappointing conclusion, we note that over short time periods network growth does appear to be approximately time independent. In many cases the origin of the time dependence may have a simple explanation, which, if accounted for in prediction models, may avoid excessive errors in forecasting the evolution of networks and the dynamics taking place on those networks. However, knowing the impact of these assumptions is only possible if simple steps are taken to check their validity. It is our hope that this paper will encourage more people to do so.

\begin{acknowledgements} M.F. thanks Renaud Lambiotte, Henrik Jeldtoft Jensen and Vaiva Vasiliauskaite for their comments. M.F. thanks the hospitality of the Miyake laboratory at Tokyo Institute of Technology where much of this work was initiated and carried out. M.F. gratefully acknowledges a PhD studentship from the Engineering and Physical Sciences Research Council through Grant No. EP/N509486/1. KO was supported by the Japan Society for the Promotion of Science (KAKENHI), Grant-in-Aid for Scientific Research (C), Grant Number 18K11500.
\end{acknowledgements}
M.F., J.L., S.A. and K.O. proposed the project and defined the k2 model. M.F. and J.L. ran simulations and analyzed the k2 model. K.O., T.S.E. and M.F. developed the mathematical background of the model. S.A., K.O. and Y.M. proposed the second degree as a network variable of interest. M.F. proposed challenging the assumptions of simple scaling in growing networks using the k2 model. J.L. analyzed the APS citation network. K.O., K.Y., Y.M., T.S.E. and K.C. supervised the project. M.F. wrote the manuscript. All authors commented on and approved the final manuscript.

\par \vspace{\baselineskip}
\setcounter{section}{0}
\setcounter{equation}{0}
\setcounter{figure}{0}
\renewcommand{\thesection}{\Alph{section}}
\renewcommand{\theequation}{\thesection\arabic{equation}}
\renewcommand{\thefigure}{\thesection\arabic{figure}}
\begin{center}
{\Large \textsc{Appendices}}
\end{center}
\section{Additional mathematical details \label{ap:maths}}
\textbf{Generalization of the k2 model.} It is possible to generalize the form of attachment shown in Eq.~\eqref{eq:kernel_simple} by including a coefficient to the second term that adjusts the total weighting of next-nearest neighbors. This can be written as
\begin{equation}
    \Pi_{i}^{(k2)} = \frac{k_{i}^{(1)} + \epsilon \sum_{\alpha = 1}^{k_{i}^{(1)}} (k_{i\alpha}^{(1)} - \delta)}{ \sum_{j=1}^{N}\left( k_{j}^{(1)} + \epsilon \sum_{\alpha = 1}^{k_{j}^{(1)}} (k_{j\alpha}^{(1)} - \delta)\right)},
    \label{eq:kernel_general}
\end{equation}
where  we require $\epsilon \geq 0$. If $\epsilon = 0$, the k2 model reduces to the BA model, $\Pi_{i}^{(k2)} \rightarrow \Pi_{i}^{(BA)}$. Alternatively, if $\epsilon = 1$, $\delta = 1$, Eq.~\eqref{eq:kernel_general} reduces to the k2 model, Eq.~\eqref{eq:kernel_simple}. In this paper, to illustrate concerns about time invariance in the scaling of attachment kernels and degree distributions, we will only focus on the $\epsilon = 1$, $\delta = 1$, case shown in Eq.~\eqref{eq:kernel_simple}. Note that the general case presented in Eq.~\eqref{eq:kernel_general} is very closely related to the 2 levels model proposed by \citeauthor{dangalchev2004} \cite{dangalchev2004}. However, the 2 levels model double counts the first degree neighbors of node $i$, $\epsilon = 1$, $\delta = 0$, and in the analysis of the model, \citeauthor{dangalchev2004} only looked at very small networks in which issues concerning the time invariance of the attachment kernel and degree distributions cannot be seen.

\textbf{Formal definition of the k2 model.} We can define $k_{i}^{(\ell)}$ as the number of unique nodes which are $\ell$ or fewer steps from the target node $i$, excluding node $i$ itself. Let $\mathcal{N}_{i\ell}(t)$ be the set of nodes which are distance $\ell$ from node $i$ in the network at time $t$, that is $\mathcal{G} (t)$ which is after all nodes and edges have been added and this has $m_0 + t \approx t$ nodes. The distance between nodes $i$ and $j$ is defined as the minimum number of edges which need to be crossed in order to form a continuous path from node $i$ to node $j$. Then we define
\begin{subequations}
\begin{align}
    q^{(\ell)}_i (t) &= | \mathcal{N}_{i \ell}(t)|,
\\
     k^{(\ell)}_i  (t) &= \sum_{j=1}^{\ell}q^{(j)}_i  (t).
\end{align}
\end{subequations}
where $k_{i}^{(1)}(t) = q^{(1)}_i(t)$. In this paper we do not consider attachment kernel's proportional to $k_{i}^{(\ell)}(t)$ for $\ell > 2$. However, it is interesting to note that if the attachment kernel were proportional to $k_{i}^{(\ell)}(t)$ and $\ell \geq D(t)$, where $D(t)$ is the network diameter, this attachment kernel is equivalent to random attachment until the growing network has diameter $D(t) > \ell$.

\textbf{Derivation of Eq.~(1).}
When $m=1$, $k_{j}^{(2)}(t)$ can be written as
\begin{equation}
k_{j}^{(2)}(t)=\sum_{\alpha=1}^{k_{j}^{(1)}(t)} k_{j_{\alpha}}^{(1)}(t).
\end{equation}
Thus, the denominator of Eq.~\eqref{eq:kernel} can be rewritten as
\begin{eqnarray}
\sum_{j=1}^{N} k_{j}^{(2)}(t)&=&\sum_{j=1}^{N} \sum_{\alpha=1}^{k_{j}^{(1)}(t)} k_{j_{\alpha}}^{(1)}(t)\nonumber \\ &=&\sum_{l=1}^{N} n_{l}(t) k_{l}^{(1)}(t)
\end{eqnarray}
where
\begin{equation}
n_{l}(t)=k_{l}^{(1)}(t).
\end{equation}
Therefore, we obtain
\begin{equation}
\sum_{j=1}^{N} k_{j}^{(2)}(t)=\sum_{l=1}^{N}\left(k_{l}^{(1)}(t)\right)^{2}.
\label{eq:denom_equal}
\end{equation}

\begin{figure}
    \centering
    \includegraphics[width=\linewidth]{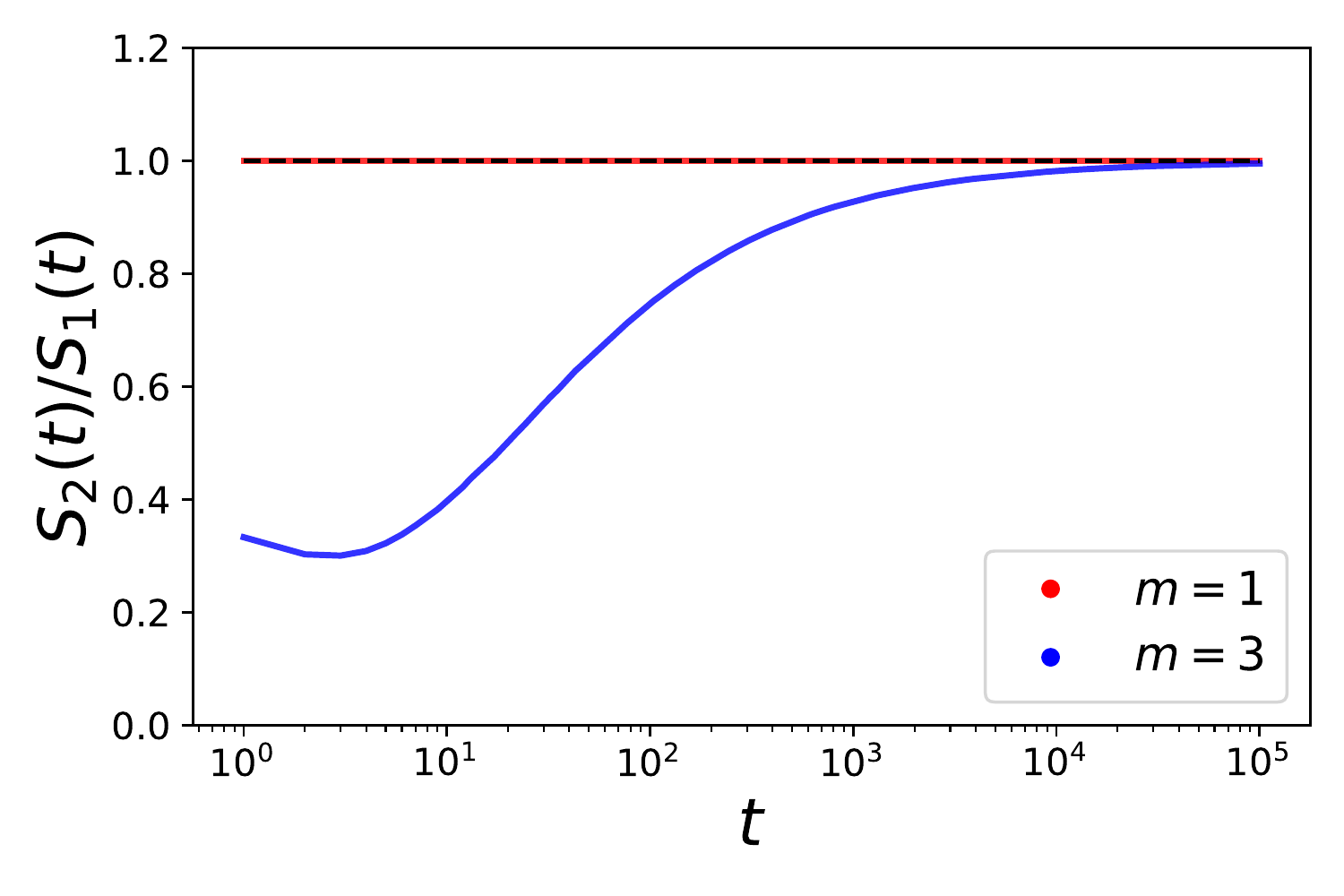}
    \caption{The ratio of the sum over the second degree of each node in the network to the sum over the first degree squared for each node indicated in the network, see Eq.~\eqref{eq:denom_ratio}. The ratio equals one for $m=1$, and converges to one for $m>1$.}
    \label{fig:denom_equal}
\end{figure}

For $m>1$, we can test the validity of Eq.~\eqref{eq:denom_equal}. Figure~\ref{fig:denom_equal} plots the ratio of the two sums, for $m=1,3$, defined as
\begin{equation}
    \frac{S_2(t)}{S_1(t)} = \frac{\sum_{j=1}^{N} k_{j}^{(2)}(t)}{\sum_{l=1}^{N} (k_{l}^{(1)}(t))^{2}},
    \label{eq:denom_ratio}
\end{equation}
against time. The figure has been averaged over 100 simulations of the k2 model. Figure~\ref{fig:denom_equal} indicates that for $m=1$, $S_2(t)/S_1(t) = 1$ for all $t$, as expected. For $m>1$, there is a noticeable difference between $S_2(t)$ and $S_1(t)$ at very small times in the network's evolution. This is to be expected since when the network is small, the probability of acquiring non-unique second degree neighbors is small, but not negligible. As the network evolves, the ratio $S_2(t)/S_1(t)$ quickly converges to $1$, with $S_2(t)/S_1(t) > 0.9$ by $t=10^3$. This indicates that Eq.~\eqref{eq:denom_ratio} is a good approximation even for $m>1$.

\textbf{Derivation of Eq.~(10).}
Let’s consider a set of nodes $i$, the neighbors $
i_{\alpha}$ connected to node $i$, and the node $i_{\alpha(\beta)}$ connected to node $
i_{\alpha}$. Here, $
\alpha=1, \cdots, k_{i}^{(1)}(t)$, and $
\beta=1, \cdots, k_{\alpha}^{(1)}(t)-1$.
Then, the k2 model with $m=1$ is applied to nodes $i_{\alpha}$ as follows:
\begin{equation}
\frac{d}{d t} k_{i_{\alpha}}^{(1)}(t)=\frac{k_{i_{\alpha}}^{(2)}(t)}{\sum_{j=1}^{N} k_{j}^{(2)}(t)} \ \text { for } \ \forall \alpha.
\label{eq:appendixb1}
\end{equation}
Next, by using Eq.~\eqref{eq:k2def}, the numerator of the right side of Eq.~\eqref{eq:appendixb1} can be rewritten as
\begin{equation}
k_{i_{\alpha}}^{(2)}(t)=k_{i}^{(1)}(t)+\sum_{\beta=1}^{k_{\alpha}^{(1)}(t)-1} k_{i_{\alpha(\beta)}}^{(1)}(t).
\end{equation}
Therefore, we obtain the dynamical equations of node $i_{\alpha}$ as follows:
\begin{equation}
\frac{d}{d t} k_{i_{\alpha}}^{(1)}(t)=\frac{k_{i}^{(1)}(t)+\sum_{\beta=1}^{k_{\alpha}^{(1)}(t)-1} k_{i_{\alpha(\beta)}}^{(1)}(t)}{\sum_{j=1}^{N}\left(k_{j}^{(1)}(t)\right)^{2}}.
\end{equation}
Summing $k_{i_{\alpha}}^{(1)}(t)$ from $\alpha=1$ to $
\alpha=k_{i}^{(1)}(t)$ for each side, we obtain
\begin{multline}
\frac{d}{d t}\left(\sum_{\alpha=1}^{k_{i}^{(1)}(t)} k_{i_{\alpha}}^{(1)}(t)\right)= \\ \frac{\sum_{\alpha=1}^{k_{i}^{(1)}(t)}\left(k_{i}^{(1)}(t)+\sum_{\beta=1}^{k_{\alpha}^{(1)}(t)-1} k_{i_{\alpha}(\beta)}^{(1)}(t)\right)}{\sum_{j=1}^{N}\left(k_{j}^{(1)}(t)\right)^{2}}.
\label{eq:appendixb2}
\end{multline}
Recall that, $k_{i}^{(2)}(t)=\sum_{\alpha=1}^{k_{i}^{(1)}} k_{i_{\alpha}}^{(1)}(t)$. Thus, Eq.~\eqref{eq:appendixb2} can be modified as follows:
\begin{eqnarray}
\frac{d}{d t} k_{i}^{(2)}(t)&=&\frac{\sum_{\alpha=1}^{k_{i}^{(1)}(t)} k_{i}^{(1)}(t)+\sum_{\alpha=1}^{k_{i}^{(1)}(t)} \sum_{\beta=1}^{k_{\alpha}^{(1)}(t)-1} k_{i_{\alpha}(\beta)}^{(1)}(t)}{\sum_{j=1}^{N}\left(k_{j}^{(1)}(t)\right)^{2}}\nonumber \\
&=&\frac{k_{i}^{(1)}(t) k_{i}^{(1)}(t)+\sum_{\alpha=1}^{k_{i}^{(1)}(t)} \sum_{\beta=1}^{k_{\alpha}^{(1)}(t)-1} k_{i_{\alpha}(\beta)}^{(1)}(t)}{\sum_{j=1}^{N}\left(k_{j}^{(1)}(t)\right)^{2}}\nonumber \\
&=&\frac{\left(k_{i}^{(1)}(t)\right)^{2}+\sum_{\alpha=1}^{k_{i}^{(1)}(t)} \sum_{\beta=1}^{k_{\alpha}^{(1)}(t)-1} k_{i_{\alpha}(\beta)}^{(1)}(t)}{\sum_{j=1}^{N}\left(k_{j}^{(1)}(t)\right)^{2}}.
\end{eqnarray}
This is equivalent to Eq.~\eqref{eq:k2_evolution}.

\section{Additional Simulation Results \label{ap:sims}}

\begin{figure}
\centering
    \subfloat{\includegraphics[width = \linewidth]{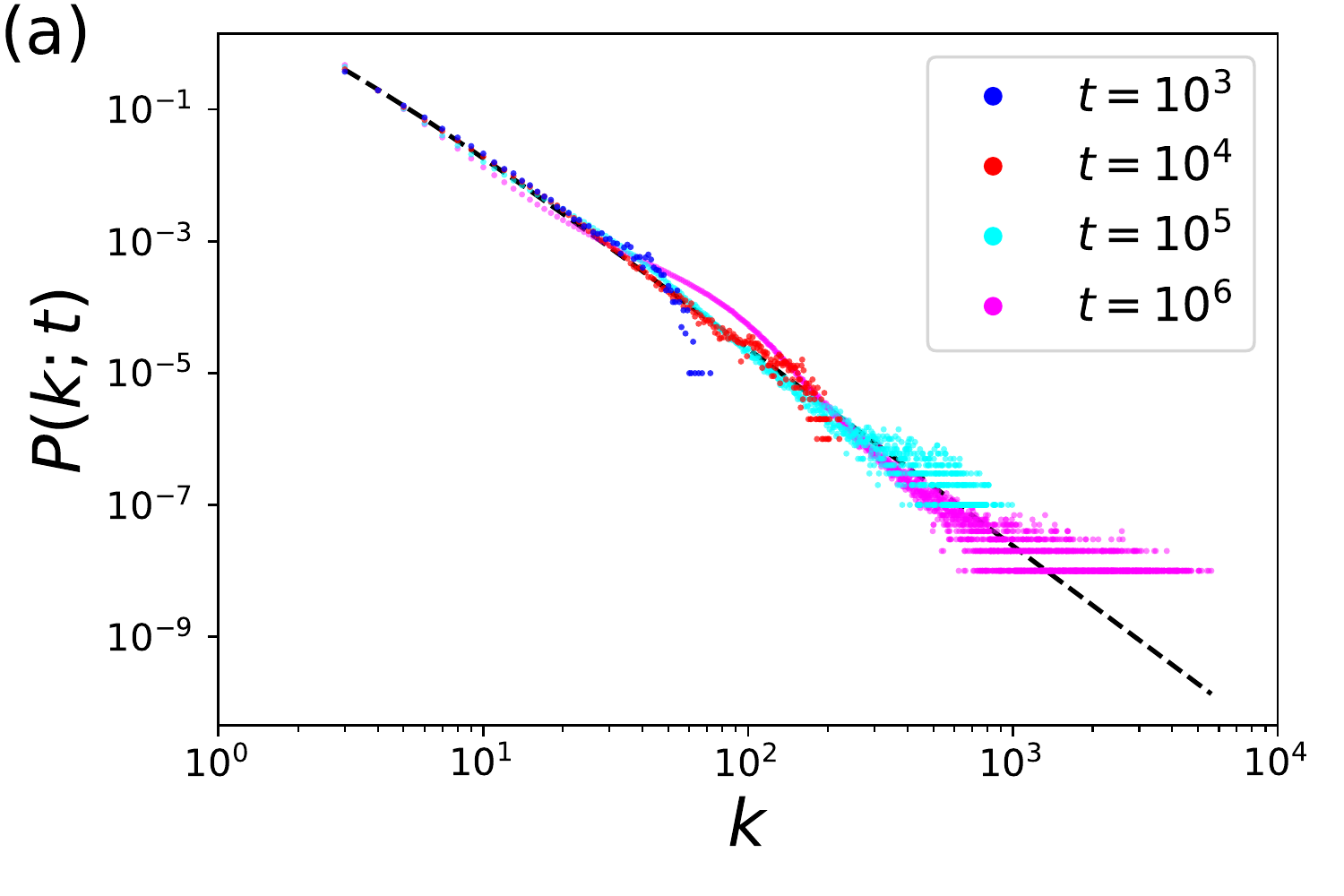}}
    \newline
    \subfloat{\includegraphics[width = \linewidth]{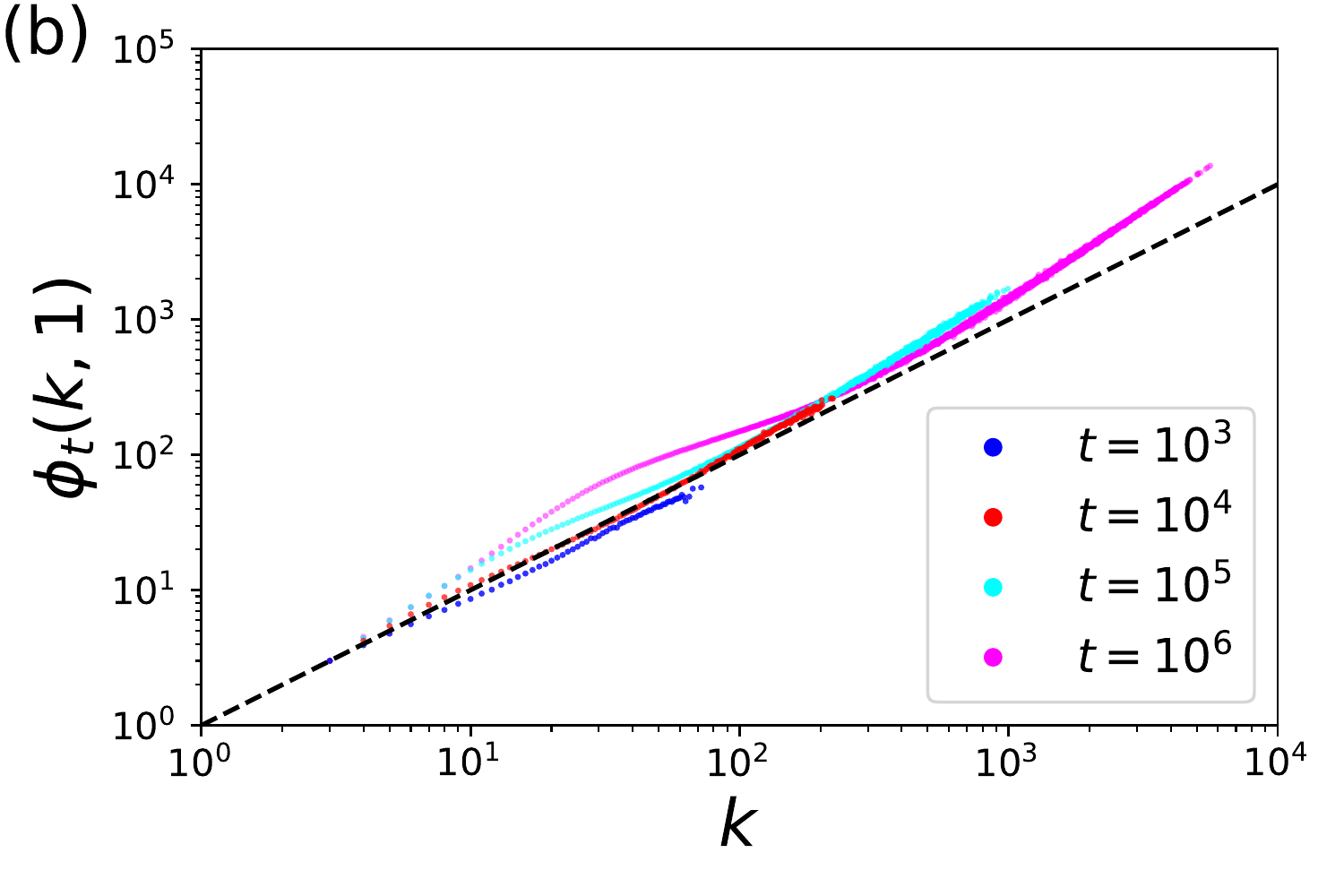}}
    \caption{The (a) degree distribution and (b) relative attachment kernel for the k2 model over time with $m=3$. The dashed lines show the expected scaling for the BA model. Early in the growth of the k2 model, the evolution of the network is largely indistinguishable from the BA model. As the k2 model grows, both the degree distribution and relative attachment kernel deviate significantly from the simple scaling predicted by the BA model.}
    \label{fig:k2_degree_m3}
\end{figure}

\textbf{Degree distribution and relative attachment kernel for $\bm{m=3}$.} In addition to the results presented in the main manuscript for the k2 model with $m=1$, Fig.~\ref{fig:k2_degree_m3} shows the degree distribution and relative attachment kernel for the k2 model with $m=3$. The results are fully consistent with those shown for $m=1$ previously. Initially, the degree distribution appears qualitatively similar to the power-law scaling expected from linear preferential attachment. This is associated with an approximately linear relative attachment kernel. However, as the network evolves, clear deviations from the simple scaling form predicted by the BA model appear in both the degree distribution and the relative attachment kernel. Note, the time for these deviations to become significant increases and $m$ is increased. The magnitude of the deviations shown for $m=1$ exceed those for $m=3$.

\begin{figure}
\centering
\subfloat{\includegraphics[width = \linewidth]{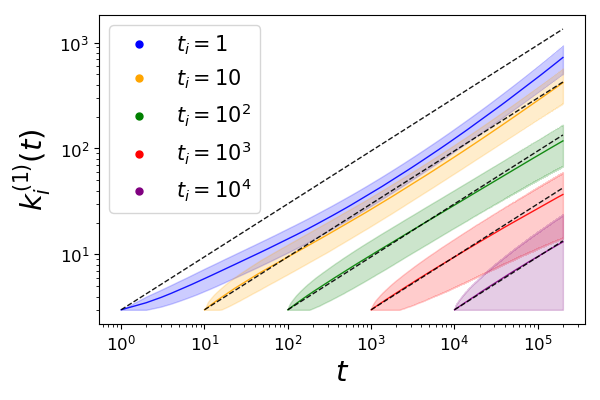}}
\caption{The evolution of the degree of individual nodes added at time $t_i$ for the k2 model with $m=3$, averaged over 10000 simulations. The shaded region around the solid lines indicates the standard deviation across the simulations. The dashed lines indicate the power function scaling that would be expected from the BA model, see Eq.~\eqref{eq:power_function}.}
\label{fig:degree_evo_m3}
\end{figure}

\textbf{Degree evolution for $\bm{m=3}.$} Figure~\ref{fig:degree_evo_m3} shows the evolution of individual nodes added at time $t_i$ in the k2 model for $m=3$. The figure is consistent with the previous result shown for $m=1$. Note in particular that the magnitude of the standard deviation is significantly smaller than for $m=1$. This suggests that the results for $m=3$ better reflect the true underlying degree scaling in the k2 model than the result for $m=1$. It is especially clear how closely nodes added at large $t_i$ follow the $t^{1/2}$ scaling predicted by the BA model during the initial phase after the node is added to the network.

Two additional details are worth highlighting: (1) After the initial transient phase during which nodes scale approximately with $t^{1/2}$, the scaling deviates from $\delta=1/2$ scaling to $\delta < 1/2$, but the magnitude of the change is much smaller than for $m=1$. This result is of particular interest since extended transient times and smaller deviations from $\delta=1/2$ scaling may explain why the transient period for the degree distribution and relative attachment kernel shown in Fig.~\ref{fig:k2_degree_m3} are longer, and follow the BA model more closely, than the equivalent for $m=1$. (2) For $t_i=10$, it appears that shortly after entering the $\delta < 1/2$ phase, the exponent increases again and appears to approach $\delta >1/2$, although the effect is very small. Longer simulations are required to clearly elucidate the scaling behavior of individual nodes, but these simulations are computationally challenging in the current framework.

\bibliography{ref}
\end{document}